\documentclass[10pt, conference, compsocconf, letterpaper]{IEEEtran}
\IEEEoverridecommandlockouts
\usepackage{bm}
\usepackage{cite}
\usepackage{array}
\usepackage{amsmath,amssymb,amsfonts}
\usepackage{algorithmic}
\usepackage{graphicx}
\usepackage{textcomp}
\usepackage{xcolor}
\usepackage{subfigure}
\usepackage{multirow}
\usepackage{ulem}
\def\BibTeX{{\rm B\kern-.05em{\sc i\kern-.025em b}\kern-.08em
    T\kern-.1667em\lower.7ex\hbox{E}\kern-.125emX}}
\begin{document}

\title{MassNet: A Deep Learning Approach for Body Weight Extraction from A Single Pressure Image
\thanks{This work was supported by the National Natural Science Foundation of China under Grant 62072420. The work involved human subjects in its research. Approval of all ethical and experiment procedures and protocols was granted by Bio-medical Ethics Committee of University of Science and Technology of China.}
}

\author{\IEEEauthorblockN{Ziyu Wu, Quan Wan, Mingjie Zhao, Yi Ke, Yiran Fang, Zhen Liang, Fangting Xie and Jingyuan Cheng}
\IEEEauthorblockA{\textit{School of Computer Science and Technology} \\
\textit{University of Science and Technology of China, Hefei, China.}
 \\
\{wzy1999, wanquan2020, zmj4527, kyrickie, fyr233, sa21011173, xieft\}@mail.ustc.edu.cn, jingyuan@ustc.edu.cn}}

\maketitle

\begin{abstract}\label{abstract}
Body weight, as an essential physiological trait, is of considerable significance in many applications like body management, rehabilitation, and drug dosing for patient-specific treatments. Previous works on the body weight estimation task are mainly vision-based, using 2D/3D, depth, or infrared images, facing problems in illumination, occlusions, and especially privacy issues. The pressure mapping mattress is a non-invasive and privacy-preserving tool to obtain the pressure distribution image over the bed surface, which strongly correlates with the body weight of the lying person. To extract the body weight from this image, we propose a deep learning-based model, including a dual-branch network to extract the deep features and pose features respectively. A contrastive learning module is also combined to the deep-feature branch to help mine the mutual factors across different postures of every single subject. The two groups of features are then concatenated for the body weight regression task. To test the model's performance over different hardware and posture settings, we create a pressure image dataset of 10 subjects and 23 postures, using a self-made pressure-sensing bedsheet. This dataset, which is made public together with this paper, together with a public dataset, are used for the validation. The results show that our model outperforms the state-of-the-art algorithms over both 2 datasets. Our research constitutes an important step toward fully automatic weight estimation in both clinical and at-home practice. Our dataset is available for research purposes at: \textit{https://github.com/USTCWzy/MassEstimation}.

\end{abstract}


\section{Introduction}\label{Intro}

Body weight, one of the most essential physiological traits, is of considerable significance in many applications like body management~\cite{naghii2006importance}, rehabilitation~\cite{duncan2011body}, nutrition management~\cite{chan1999nutrition}, and drug dosing\cite{ekwaru2014importance} for patient-specific treatments. Considering the close relationship between human body weight and physiological status, long-term and routine weight records shall be important for not only personal health management and medical care, but also for potential medical knowledge. 

Traditional methods like spring scales, weight measuring beds, anthropometric approximation, or direct observation have their shortcomings like limited subject selection (not suitable for disabled persons), cost, or inaccuracy. Vision-based methods developed quickly in recent years and have been applied to estimate BMI. They however face privacy issues, especially in hospitals.

Our work is motivated by the fact that the body weight is always ''printed'' on the surface the human contact. Especially, the bed would be a good choice because everyone lies on it every day and during the 8-hours sleep, the whole body weight is on the bed. A pressure sensing matrix, built into a bedsheet or mattress, could be a non-invasive and privacy-preserving tool to collect such ''pressure images''~(Typical pressure images can be found in Fig.~\ref{fig:example_heatmap}). Such film-based or textile-based pressure mapping tools have already been developed and applications like pose estimation~\cite{luo2021intelligent}, 3D shape estimation~\cite{clever2020bodies}, exercise quality evaluation~\cite{zhou2022quali} have been explored. Compared to traditional commercial or clinical devices, a pressure-sensing mattress can be deployed to existing environments non-intrusively and provide a sensitive but stable approach to monitor and analyze users' long-term behavior information~(e.g. postures and activities), biomedical factors~(e.g. body shape) and physical conditions~(e.g. skin conditions and sleep quality) without disrupting their daily life. Extracting accurate body weights from collected pressure images will effectively serve users' nutrition management and disease prevention.

While data acquisition technologies for acquiring pressure images are there already, to the best of my knowledge, there is yet no specially-developed algorithm for extracting body weight from the pressure images. Although directly summing up the weight distribution seems a solution, it performs poorly in reality~(we will show this in Section~\ref{overall}), due to the non-ideal sensor characteristics such as hysteresis, non-consistency among sensors, and sensor deformation during long-term usage. The closest work is Body Mass Index~(BMI) extraction from pressure images~\cite{davoodnia2020deep}. However, body weight gives a more intuitive perspective on medical services. Converting BMI to body weight requires the subject to hold a certain posture~(usually straightening) to estimate the height, performing poorly for the elderly and the disabled. The body weight is also of a broader distribution~(from 3 $kg$ for the newborns to $>100kg$ for the adults) than the BMI~(from 18.5 to 24.9 $kg/m^2)$, all for healthy people. The state-of-the-art deep learning-based feature extractors don't take pressure images' attributes into account and often need a large dataset for training, which is another bottleneck for pressure images. Their performance in the body weight extraction task is still unknown and doubtful.

To extract body weight from pressure images requires thus the answers to the following three questions: 

\begin{enumerate}
\item How to customize the neural network, taking into consideration the nature of pressure images?
\item Are the methods valid on different pressure image datasets, corresponding to different hardware and/or posture settings?
\item What is the best accuracy of the extracted weight, using all feasible methods (viz. the proposed network, the feature-based regression, and pre-trained classic neural networks) and under different model settings?

\end{enumerate}

For the above purpose, we propose a dual-branch deep learning-based model to extract the body weight from a single pressure image. Specifically, the model comprises a deep feature extractor as the main branch and a multiple-layer perception~(MLP) as the auxiliary joint branch to extract the deep features and joint features, respectively. A refined supervised contrastive loss is exerted on these deep features to learn the distributions of multiple posture images and explore the posture-invariant and mass-related features after projecting the features into high-level latent space. Finally, both features are concatenated in the data fusion module to estimate the body weight. We validate the proposed model on two datasets, namely: SLP dataset~\cite{liu2020simultaneously} using a Tekscan film pressure mat and a self-created dataset created using a self-developed pressure mapping bedsheet.

The contributions can be summarized as follow:
\begin{enumerate}
    \item We developed a dual-branch deep learning-based model proposed for estimating human body weight from a single pressure image. A refined supervised contrastive loss is used to extract the mutual high-level features across different postures of every single subject. To the best of our knowledge, our work is the first attempt to extract body weight directly from pressure images and also the first work using contrastive learning techniques to help estimate human biomedical factors.
    \item We validate our model on two datasets, that differ from each other in the sensor matrix and posture settings. This includes a self-created dataset using a pure-textile sensing matrix, containing 10 subjects~(7M, 3F), 23 postures, and a total of 918 pressure images. The dataset is made public, well annotated, and can be used also for classification, joint estimation, and other tasks. 
    \item Our model outperforms the state-of-the-art models on both datasets, with a minimal $4.59kg$
    prediction error on the SLP dataset and $1.5kg$ on our collected dataset.
        
\end{enumerate}

The rest of the paper is organized as follows. Section~\ref{related_work} introduces the related works. Section~\ref{sec:data} introduces the sensing matrices and the datasets. Section~\ref{models} describes the details of our proposed network, and Section~\ref{experiments_results} presents our model's performances on three datasets, compared with the results from featured-based regression and other pre-trained classic neural networks, followed by the detailed analyses with ablation studies. Finally, Section~\ref{conclusions} concludes the paper.

\section{Related Work}\label{related_work}

\subsection{Body Weight Estimation}
The most traditional way to measure the body weight, viz. using a scale, is of low cost and accurate. However, it requires the subject to stand onto the scale, unfriendly to those who have difficulties in moving. Experienced physicians can estimate the body weight just by observing the body shape, which is a quick and cost-free method, however, has been proven inaccurate~\cite{menon2005accurate}. Anthropometric approximation obtains the body weight by measuring specific parameters like body height, and waist and hip circumference then deducing the body weight with a linear regression model. It is an accurate method~\cite{lorenz2007anthropometric} but requires time to measure and is also unfriendly to patients that can not move.

Vision-based machine learning approaches provide a fully-automatic and contactless way to monitor human body weight or BMI. A commonly used pipeline for vision-based weight estimation approaches includes human segmentation from the background, feature extraction, and weight regression. Wen et al.~\cite{wen2013computational} firstly estimated BMI with facial images. Nguyen et al.~\cite{nguyen2014seeing} used RGB-D images to predict body weights by training SVR with biomedical traits and extracted sideview shape features. Jiang et al.~\cite{jiang2019body} proposed a body weight analysis framework to estimate the BMI value from a single image. They employed a body contour and skeleton joints detection module to obtain the human silhouette and calculate its BMI with 5 curated anthropometric features. Huang et al.~\cite{huang2021seeing} expanded another two anthropometric features and proposed a 4-branch network for the BMI estimation task by extracting 3D, deep, statistics, and anthropometric features from a single image, respectively. Attention is also applied for BMI estimation after removing irrelevant background in~\cite{jin2022attention}. However, all these works require human bodies or faces facing the camera, which is unrealistic in many situations. In that case, Altinigne et al.~\cite{altinigne2020height} proposed a multi-task U-Net-based network to predict human attributes. This method allows arbitrary poses but performs poorly, with a $9.8kg$ mean absolute loss. Other researchers also focus on the possible occlusion issue. Alexander et al.~\cite{bigalke2021seeing} studied the weight estimation of covered patients. They implemented a 3D U-Net to construct the 3D points clouds of the subject under blankets, and then a 3D CNN was applied for the weight regression. In summary, vision-based weight estimation methods have proven efficient with 2D/3D body and facial images. However, camera angles and occlusions will seriously affect their performances. Besides, illumination and privacy issues also restrict their application environment.

To sum up,  purchasing or updating equipment ~(e.g. bed scales) is one of the most precise weight acquisition methods but needs users to weigh actively or passively, facing financial and physical pressure and disturbance to daily life. Estimating weight from anthropometric information is convenient but leads to errors and is impossible for long-term monitoring. Vision-based approaches provide fully-automatic workflows but are confined by user postures, dressing, illumination, and privacy issue. Pressure mattress has a strong potential in human biomedical traits extraction, but little attention has been paid to this area. We found only few previous works focusing on body weight estimation with pressure images including~\cite{davoodnia2020deep} with limited and onefold features for BMI calculation and~\cite{kim2020prediction} with 128 sensing units and only 60 samples. Furthermore, a customized and pressure images nature-based neural network architecture is still missing. Thus, we propose the MassNet, a dual-branch network to utilize the pressure image information and joint features to predict accurate body weights.

\subsection{Contrastive Learning}

Contrastive learning is a self-supervised learning approach using triple loss~\cite{weinberger2009distance}, contrastive loss~\cite{hadsell2006dimensionality}, or InfoNCE loss~\cite{oord2018representation} to project features extracted into a latent space, in which positive pairs are close to each other, whereas negative samples are far apart. Contrastive learning has achieved noticeable performance in many downstream tasks such as human activity recognition and biosignal process~\cite{haresamudram2021contrastive,banville2021uncovering}. For example, DELDARI et al.~\cite{deldari2022cocoa} proposed a cross-modality contrastive learning~(COCOA) framework by calculating the cross-correlation between latent encodings of different modalities of an individual sample and maximizing inter-modality agreements. Shen et al.~\cite{shen2022contrastive} assume that the neural activities of the subjects are in a similar state when they receive the same segment of emotional stimuli and use the InfoNCE loss to solve the inter-subject variability of emotion-related EEG signals issue. Zhang et al.~\cite{zhang2021dynamic} use an auxiliary contrastive loss that digests the object-related embeddings to improve their model's generalization ability to unseen subjects. Inspired by their work, we integrated contrastive loss into our network to support extracting the pose-irrelevant and subject-specific features related to the weight estimation task.

\section{Data Description}\label{sec:data}
Before moving to the algorithm design, we first introduce here the datasets and the sensing mat, which serve as the foundation of the algorithm. 

\subsection{The Sensing Principle}
\label{sec:datadescrp}

Pressure distribution over a surface can be converted into a pressure image with a pressure-sensitive mat. Such mat is normally composed of hundreds or thousands of pressure sensors,  organized in a matrix format. A typical resistive pressure mat is made of 3 layers: The upper and lower layers contain multiple parallel conductive paths, put to 90 degrees and separated by a middle resistance layer, whose resistance changes under the physical stimulus. Each crosspoint of two conductive paths constitutes a sensing unit, whose resistance represents the local pressure. The matrix is driven by IOs with specific voltage protocols on one side and the voltages are read out on the other side, corresponding to the sensing units' resistance. The detailed driving architectures and the resistance-voltage mapping scheme can be found in~\cite{Shu2015New, Bo2014From}

%

\begin{figure}[t]
\centering
    \subfigure[SLP]{\includegraphics[width=0.14\textwidth, height=4cm]{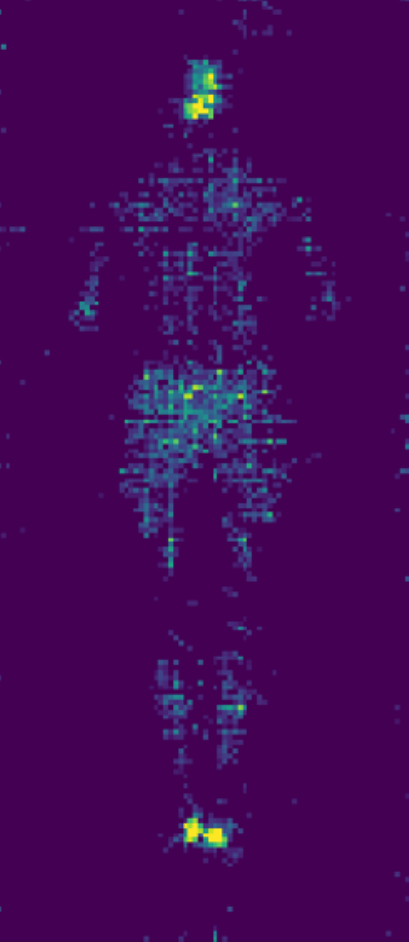}\label{fig:example_heatmap_slp}}
    \subfigure[Ours]{\includegraphics[width=0.14\textwidth, height=4cm]{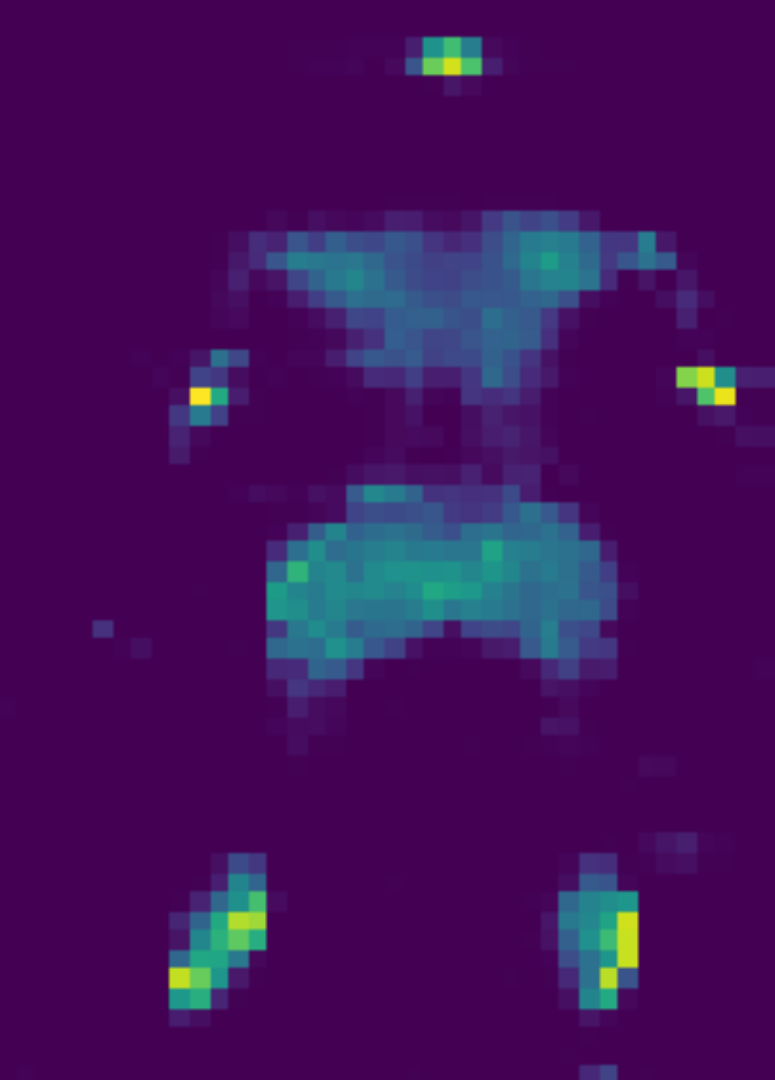}\label{fig:example_heatmap_ours}}
    \subfigure[HRL-ROS]{\includegraphics[width=0.14\textwidth, height=4cm]{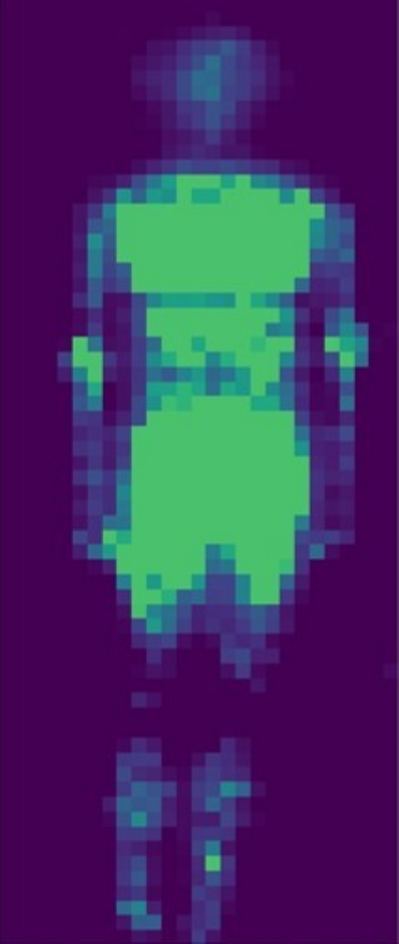}\label{fig:example_heatmap_hrl}}
\caption{Example pressure images of the supine pose from 2 public datasets and the self-created dataset}
\label{fig:example_heatmap}
\end{figure}

When such a matrix is made into a bedsheet or mattress, the pressure ''image'' of the lying person can be obtained, where the value of each sensing unit is defined as a ''pixel''. Fig.~\ref{fig:example_heatmap} demonstrates the supine pose from 3 datasets. It can be seen that although the human shape and the posture can be easily recognized from all the 3 images, they differ from each other in both the spatial pressure distribution and the data dynamic range. For example, in Fig.~\ref{fig:example_heatmap_slp}, the highest pressure is on the head and heals, and in Fig.~\ref{fig:example_heatmap_ours} demonstrates a high pressure also on the shoulder and the sacrococcygeal regions.  In Fig.~\ref{fig:example_heatmap_hrl} the whole back and buttocks region are completely over-saturated. Fig.~\ref{fig:example_heatmap_slp} owns the highest spatial resolution but the pressure distribution's spatial continuity is not assured. These differences are a combined result from the sensor mat characteristics~(including both the sensing material and the driving circuit), the environment~(whether there are other layers of hard or soft, thick or thin sheets or mattresses above or below the sensing mat), and the test subject~(postures and weights). Noise might also appear due to the deformation of the sensor mat~(fold or other deformations randomly appearing during usage). All these non-ideal phenomena in the data create challenges in the weight estimation task.

\subsection{Datasets}

To train and evaluate the effectiveness of the proposed body weight estimation network, we need a dataset. As an emerging field, only few pressure image datasets are made public. Luckily, the majority of these have focused on lying postures. They are: PMat dataset~(2017)~\cite{pouyan2017pressure}, HRL-ROS~(2018)~\cite{clever20183d}, SLP dataset~(2020)~\cite{liu2020simultaneously}, and BodyPressure Synthetic dataset~(2020)~\cite{clever2020bodies}.

For this study, we discard the PMat dataset for its large number of duplicate data frames and HRL-ROS dataset for duplicate frames and over-saturation phenomenon. We also put the BodyPressure synthetic dataset~\cite{clever2020bodies} aside. Actually, we have started with this dataset because it is a synthetic dataset and contains the largest number of frames. But the networks trained on this dataset perform badly on all the other datasets. We later also discovered the networks trained on one dataset perform on general badly on other datasets. We believe this is mainly caused by the different sensor and environment settings in different datasets. As the BodyPressure synthetic dataset is a simulated dataset, whose sensor characteristics might match no real sensor matrix, and we are not considering domain adaptation in this paper, we discard this dataset. 

Thus only one public dataset~(SLP) is used, where the sensing mat is a film mat, which might be uncomfortable for long-term monitoring due to its airtightness. To further enhance the richness of the dataset, we created a dataset using the textile pressure mat made in our lab. These datasets are described below. Table~\ref{tab:dataset_comparision} provides the overview of all datasets.

\begin{figure}[t]
\centering
    \subfigure[Experiment setup]{\includegraphics[width=0.23\textwidth, height=4.5cm]{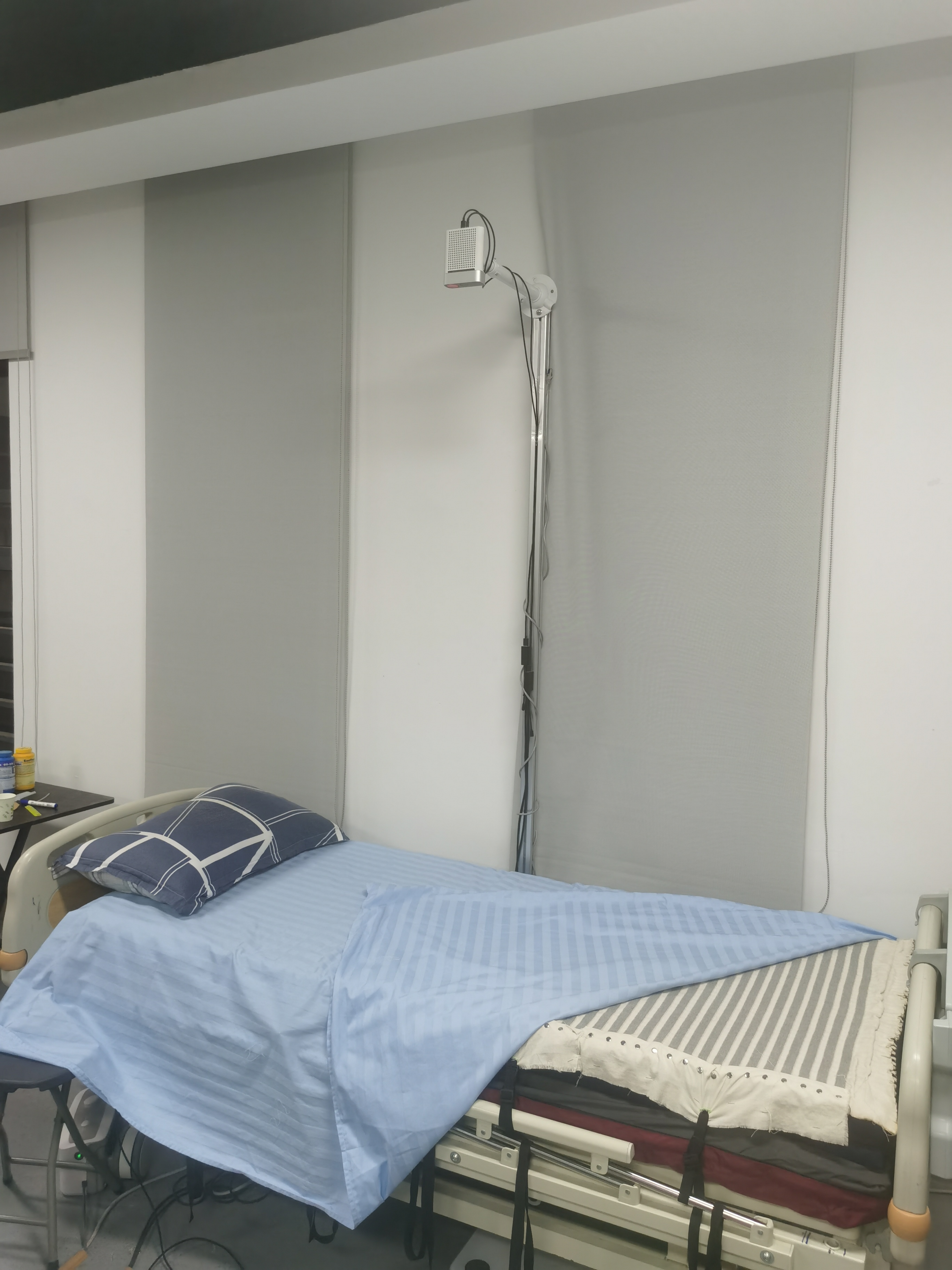}\label{fig:our_experiment_setup_and_fabric_data_setup}}
    \subfigure[The pressure sensitive fabric]{\includegraphics[width=0.23\textwidth, height=4.5cm]{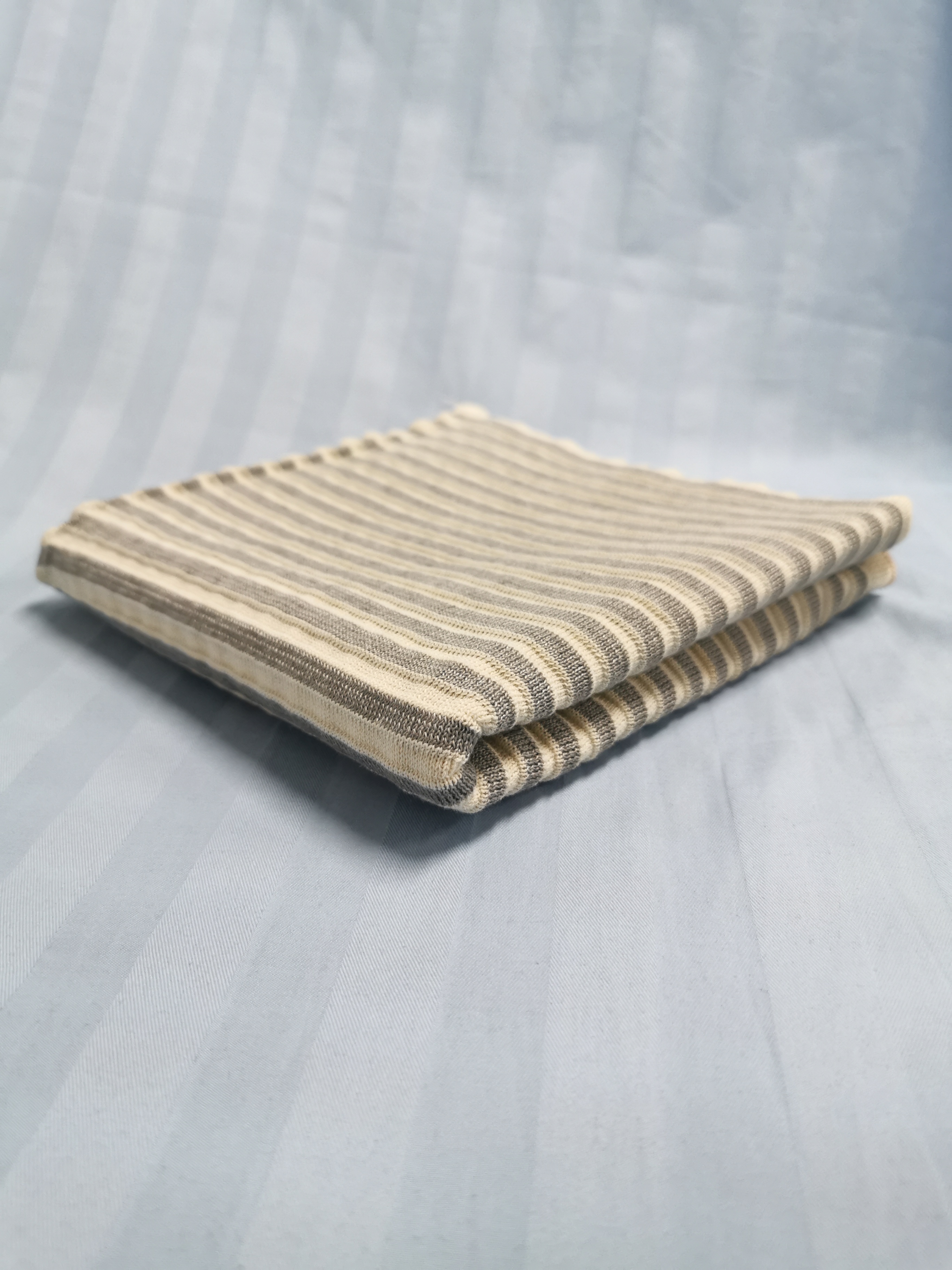}\label{fig:our_experiment_setup_and_fabric_data_fabric}}
    \subfigure[A subject in the supine pose]{\includegraphics[width=0.23\textwidth, height=4.5cm]{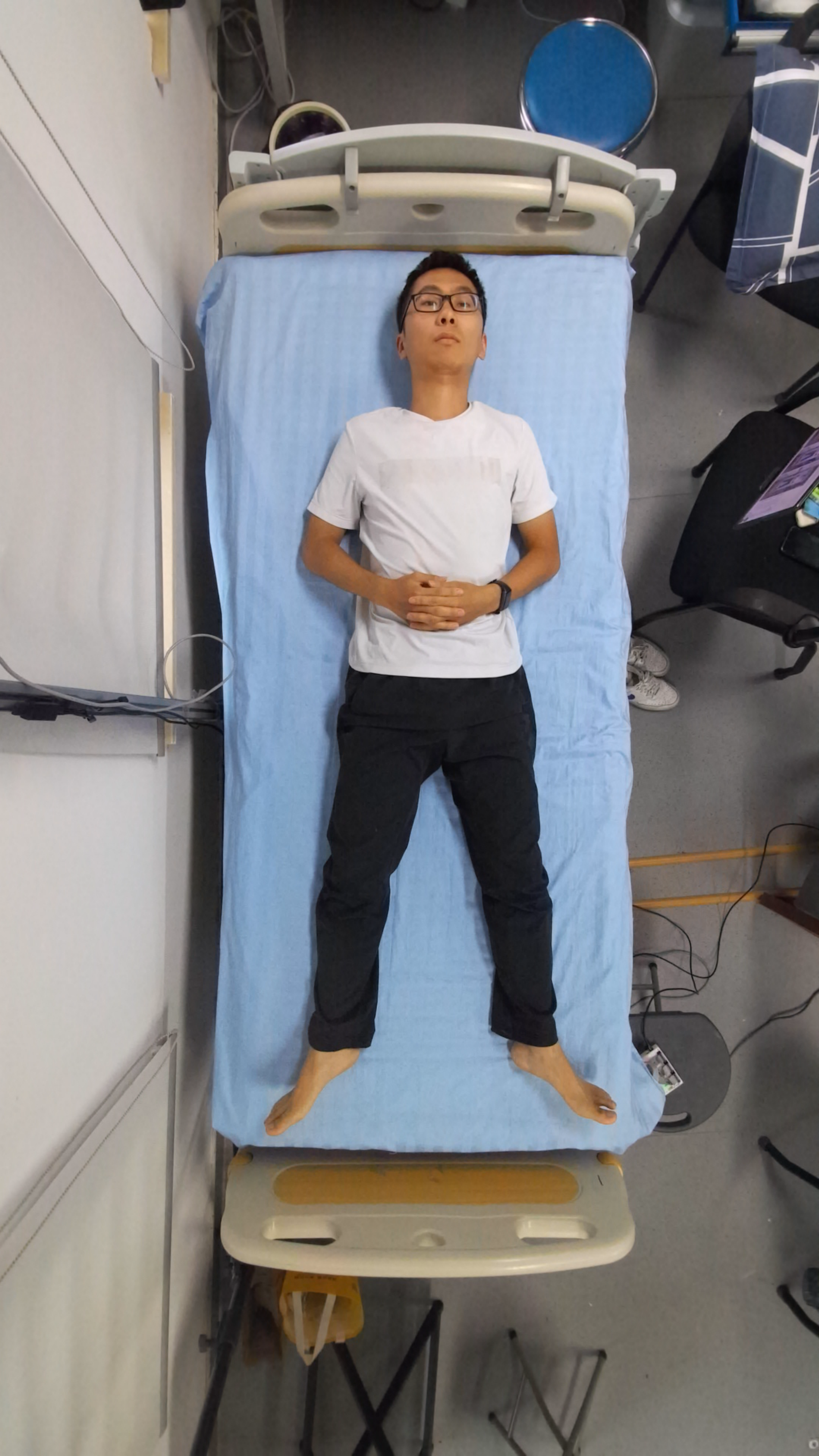}\label{fig:our_experiment_setup_and_fabric_data_image}}
    \subfigure[pressure image]{\includegraphics[width=0.23\textwidth, height=4.5cm]{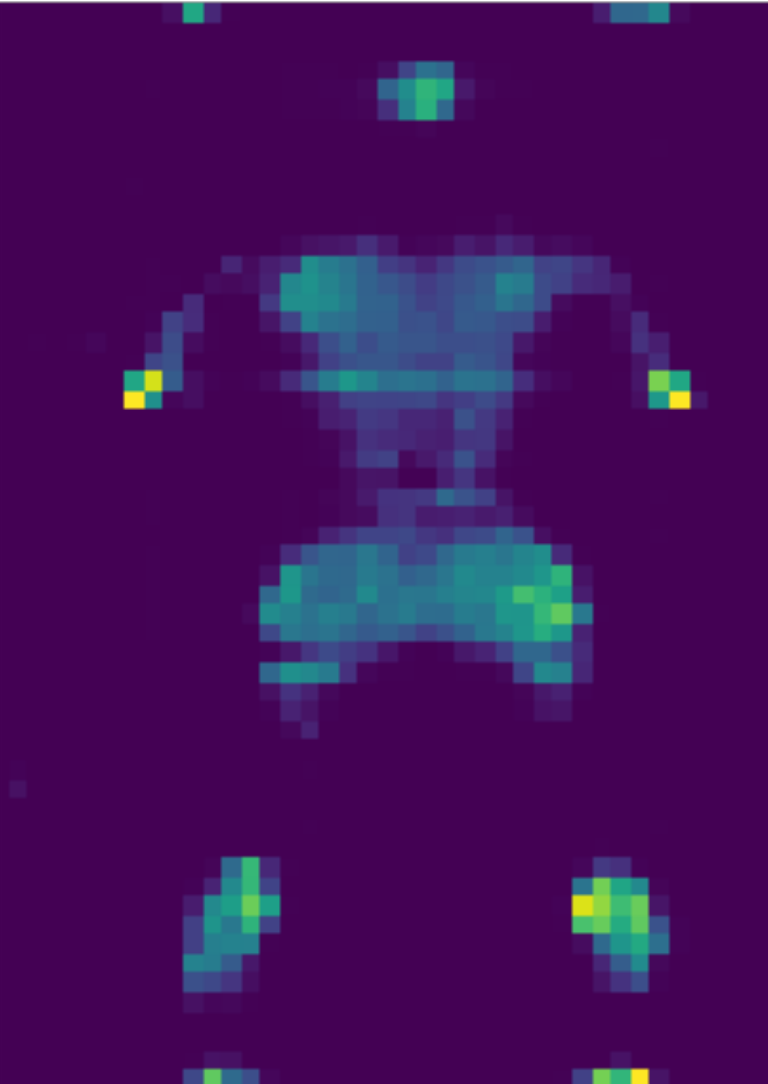}\label{fig:our_experiment_setup_and_fabric_data_data}}
\caption{Experiment setup and a representing pressure image of our datasets}
\end{figure}

\begin{table}[t]
\vspace{-0.25cm}
\caption{Datasets Overview}
\begin{center}
\setlength{\tabcolsep}{3pt}
\begin{tabular}{p{2.5cm}<{\centering}p{1.6cm}<{\centering}p{1.2cm}<{\centering}p{1.2cm}<{\centering}p{1cm}}
\hline
\textbf{Dataset} & \textbf{weight range ($kg$)} & \textbf{No. of frames} & \textbf{No. of pixels} & \textbf{type}  \\
\hline
SLP& $44.55$-$105.1$& $4590$&$192$ $\times$ $84$ & film\\
MassNet~(static)& $38.9$-$80.8$& $918$&$56$ $\times$ $40$ & fabric\\
MassNet~(dynamic)& $38.9$-$80.8$& $368313$&$56$ $\times$ $40$ & fabric\\
\hline
\end{tabular}
\label{tab:dataset_comparision}
\end{center}
\end{table}

\textbf{SLP Dataset}: It includes two environment settings (the laboratory setting and simulated hospital setting) with four modalities, specifically, RGB, Long Wavelength Infra-Red, depth, and pressure images~(collected using a Tekscan BRE5315-8 commercial film Body Pressure Measurement System~(BPMS), sensing area $1.95 \times 0.85cm^2$, with $192 \times 84$ sensing units, viz. a pitch of $1cm \times 1cm$). Only the pressure images in the laboratory environment are used in this paper for adequate participants. 102 subjects~(74 males and 28 females, height: $1.48$-$1.84m$, weight: $44.55$-$105.1kg$) participated in the experiment of lying on the mattress naturally in 45 different poses. The dataset contains 4590 images for every modality. All images are carefully calibrated with 2D keypoint annotations, body weight, height, and other physiological factors. 


\textbf{MassNet Dataset~(ours)}: We also created an own bedsheet with a full-textile sensor matrix developed by our lab, which is specially designed for long-term usage~(e.g. using stainless steel blended yarn instead of silver coated fiber for washability, double-layered jacquard knitting structure for stretchability and twistability). The experiment setup is shown in Fig.~\ref{fig:our_experiment_setup_and_fabric_data_setup}, including a hospital bed, an Azure Kinect RGBD Camera, and the pressure mapping bedsheet~($1.96 \times 0.96m^2$, with $ 56 \times 40$ sensing units and a pitch of $3.1cm \times 2.4cm$. Ten subjects took part in the experiment~(height: $1.52$-$1.73m$, weight: $38.9$-$80.8kg$, age: $22$-$27$), performed 23 different poses randomly for 4 rounds to collect the static pose dataset and repeated 14 continuous poses 10 times to collect the dynamic dataset. All poses specified are under three main categories of supine, left or right side, and prone. The whole dataset contains 918 static pose pressure images and 368313 dynamic pressure images. The study was approved by the institutional review board (IRB) of USTC.

\section{MassNet: the model structure}\label{models}


\begin{figure*}[htbp]
\vspace{-0.5cm}
\setlength{\abovecaptionskip}{-0.05cm}
\centering
   \subfigure[The architecture of MassNet]{\includegraphics[width=\textwidth, height=5cm]{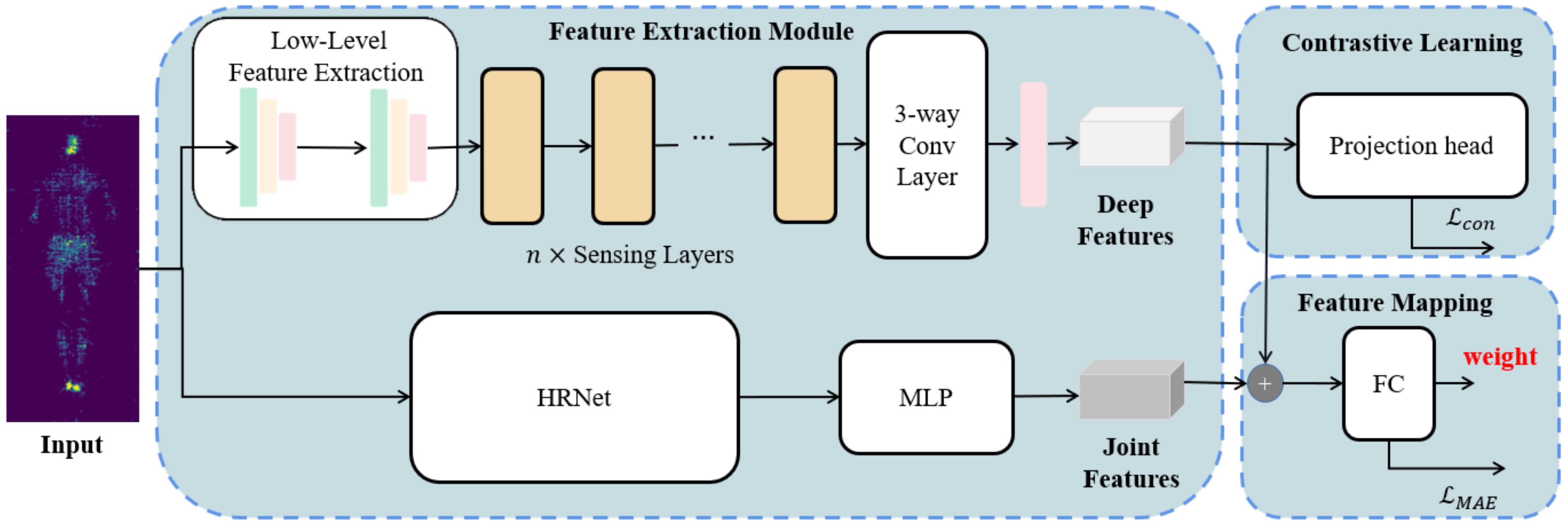}\label{fig:network_framework}}
   \subfigure[The architecture of sensing layers]{\includegraphics[width=\textwidth, height=4.35cm]{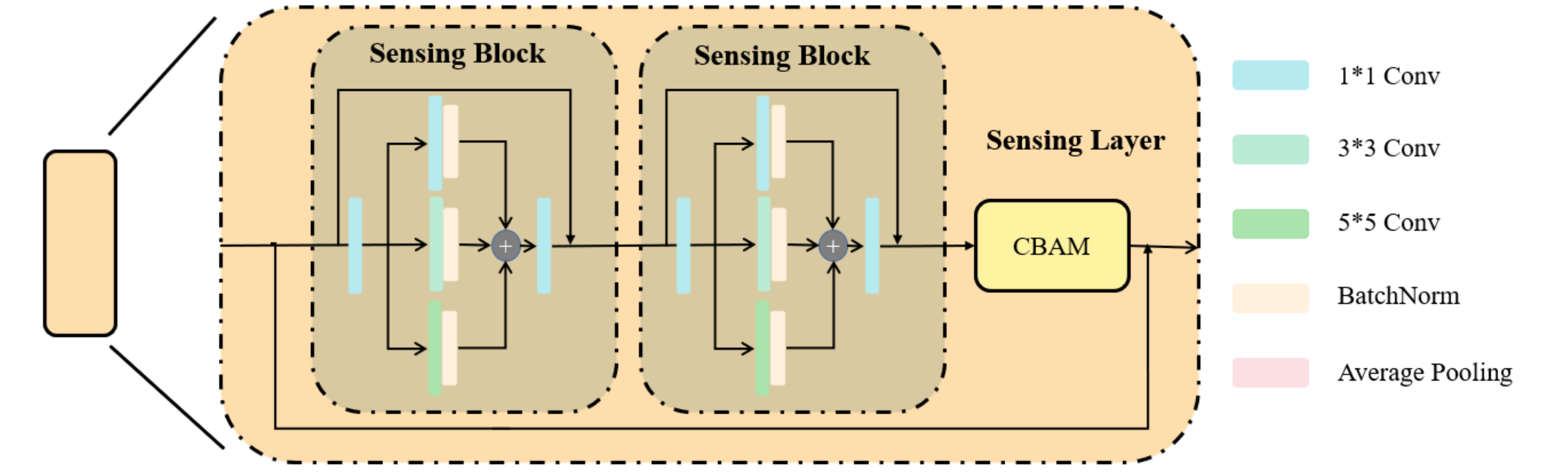}\label{fig:network_block}}
\caption{The structure of our proposed model MassNet}
\label{fig:network_architecture}
\end{figure*}

To fully integrate the nature of pressure images into the network structure, we design a dual-branch deep neural network to infer the body weights, which mainly includes 1) a feature extraction module containing stacked 3-way parallel convolution layers with different kernel sizes to understand the pressure distribution mode in different perceptual scales, 2) a contrastive constraints module to digest the posture-invariant and weight-related features, 3) a joint feature extraction branch to supply auxiliary posture information and 4) a feature mapping module to concatenate features and regress the predictions. The model structure is presented in Fig.~\ref{fig:network_architecture}, and the component details of the proposed are justified below.

\subsection{Deep Feature Extraction}

The deep body weight feature extraction module has a multi-stage design by stacking multiple sensing layers, which consists of 2 sensing blocks and a Convolutional Block Attention Module~(CBAM)~\cite{woo2018cbam} behind them to emphasize meaningful features along the channel and spatial axes. The sensing block uses a bottleneck architecture like ResNet~\cite{he2016deep}, which comprises two 1$\times$1 convolutional layers to reduce the parameters and one layer to extract features. Besides, considering that the data collected from the mattress are all centimeter-scale pressure distribution images, the local pressure values are not only correlated with their surroundings~(e.g., thicker limbs usually lead to larger contact areas and higher peaks) but also with the global pressure distribution (e.g., the pressure distribution is affected by weight, height, postures, limb length, body proportion). Therefore, multi-scale perceptual fields are needed to grasp better the connection between the local and global pressure distribution. Inspired by the design principles from Inception V1~\cite{szegedy2015going}, we use a 3-way parallel convolution layer with kernel sizes 1$\times$1, 3$\times$3, and 5$\times$5 instead of the commonly used single way 3$\times$3 convolutional layers, to give the network a wider reception field to learn the pressure distribution mode. Each 3-way convolution layer in the sensing block is followed by a batch normalization layer (ReLU Layers are not applied for the conclusions raised by MobilenetV2~\cite{sandler2018mobilenetv2}). Shortcut connections are also applied to speed up convergence in the sensing blocks and layers.

Besides the stacked sensing layers, the deep feature extraction module also has two Conv-BatchNorm-LeakyReLu structures in the front of the module to extract low-level feature maps from the original pressure images and a 3-way convolution layer with an Average Pooling layer after the sensing layers to generate the feature vectors for following feature map and contrastive constraints module.

\subsection{Contrastive Constraints Module}

Body weight is an essential physiological fator and normally remains nearly constant for a short time. Therefore, an intuition here is that under the weight prediction task, the feature vectors extracted from the pressure images of the same subject should be close in the embedded high-level space. Inspired by the SupCon~\cite{khosla2020supervised}, we introduce the supervised contrastive loss as an auxiliary loss to help form a high-cohesion and low-coupling embedding space. The deep features produced by the main branch from the same subject in a minibatch are considered as the positive pairs and meanwhile features from different subjects as the negative pairs, encouraging the network to learn the posture-invariant attributes. Within a multi-viewed batch, let $N$ be the batch size and $i \in I \equiv  \{1 ... 2N\}$ be the index of a random augmented sample. We can then define the auxiliary contrastive loss as follows:

\begin{equation}
    \mathcal{L}_{con} = \sum_{i \in I}\frac{-1}{|P(i)|}\{\sum_{p \in P(i)} log\frac{exp(m_i \cdot m_p / \tau)}{ \sum_{a \in A(i)} exp(m_i \cdot m_a / \tau)} \}
\end{equation}

where $m_i$ is the embedding of sample $i$, $\tau \in \mathcal{R}^+$ is a scalar temperature factor, $P(i)$ is the set of indices of all positive pairs in the multi-viewed batch distinct from $i$, $|P(i)|$ is its cardinality, and $A(i) \equiv I \setminus \{i\}$ is the indices set without the anchor $i$.

Although the auxiliary contrastive loss could effectively help learn the common attributes between various posture images from the same subject, it does not work well in improving prediction precision. The reason behind might be that we should focus on the subject's body weight instead of the identity. Features from two subjects with closer weights should also be adjacent in the high-dimension embedded space. While the embedded features of two subjects whose body weights are far from each other should also be far apart. We noticed that the supervised contrastive loss ignores the inter-class connections between different subjects. Thus a penalty factor $\delta$ is introduced when calculating the similarities between two features to represent the association between different subjects' physiological information. Two conditions should be met in designing this penalty factor: 1) It should reflect the body weight gap between two subjects, increase and decrease according to the body weight difference. 2) It should be $1$ when two samples are from the same subject. We thus define the penalty factor $\delta_{ij}$ as:

\begin{equation}
    \delta_{ij} = exp^{|M_i - M_j| / M_i}  
    \label{eq:deltaij}
\end{equation}

Here, $i, j \in I$ and $M_i$ means the label~(body weight) of the sample $i$. The contrastive loss then becomes: 

\begin{equation}
    \mathcal{L}_{con} = \sum_{i \in I}\frac{-1}{|P(i)|}\{\sum_{p \in P(i)} log\frac{\exp(\delta_{ip} \cdot m_i \cdot m_p / \tau)}{ \sum_{a \in A(i)}   \exp(\delta_{ia} \cdot m_i \cdot m_a / \tau)} \}
\end{equation}

Projecting feature vectors into a latent space before applying the contrastive loss has been widely adopted, whose efficacy has been proved in SimCLR~\cite{chen2020simple}.  We adopt this approach and use a 2-layer MLP as the projection head in the contrastive constraints module, enabling us to measure the distances in the projection space.

\subsection{Joint Feature Extraction}\label{chap:joint_fea_ex}

The pressure image changes with the human posture. Different postures may lead to different contact modes and pressure distribution patterns. Though the contrastive learning module could help digest the posture-invariant features, constraints from the sample size and posture diversities restrict its learning potential. To compensate for the limitation and improve the generalizability for unseen subjects, we utilize the joint positions to supplement the posture information. To rephrase it, assuming that postures can help provide a unique perspective for body weight prediction tasks, we design the joint feature extraction module as an auxiliary branch, which uses a 3-layer multilayer perceptron~(MLP) to encode the 2D joint position matrix on a pressure image into a 128-d feature vector to capture the posture information. Experiments in Section~\ref{ablation_study_conloss} verify that the collaboration between posture-invariant features and joint features enhances the performance effectively.

Currently, the 2D/3D keypoint annotations of pressure images mainly rely on their corresponding RGB images and the classical RGB-based pose estimation models~(e.g. OpenPose): RGB images as input, and the output joint locations are projected into the pressure map coordinates. Nevertheless, in many scenarios, privacy policy and adverse vision conditions like occlusions will block the acquisition of RGB images and hinder the pose estimation model's precisions. On that account, to improve the generalization of our method in different circumstances, we also try to directly obtain the join positions from the pressure image, using the HRNet~\cite{sun2019deep}, which is a state-of-the-art model in the field of human pose estimation and has shown superior performance in the pose estimation task for pressure images in~\cite{liu2020simultaneously}. In Section~\ref{joint_accuracy} we demonstrate that this module will allow our model to fit in more complicated scenarios with little precision loss.

To sum up, the whole pipeline of this branch is to take the pressure images as input, generate the joint locations with the HRnet, then extract the joint features with a 3-layer MLP for the follow-up prediction module.

\subsection{Feature Mapping}

Body weight estimation can be defined as a regression problem that maps the features extracted to the body weight. In this work, We apply a 1-layer fully-connected layer to train the regression function. The deep feature vector and the joint feature vector are concatenated as the input and Mean Absolute Error~(MAE) loss is used as the loss function, noted as $L_{MAE}$, to measure the L1 distance between predictions and the physical measured ground truth for back propagation.

\subsection{Overall Loss}

The overall loss is then a weighted sum of body weight prediction MAE loss $\mathcal{L}_{MAE}$ and the auxiliary contrastive loss $\mathcal{L}_{con}$ with a weight parameter $\lambda$:
\begin{equation}
    \mathcal{L}_{all} = \mathcal{L}_{MAE} + \lambda \mathcal{L}_{con}
\end{equation}

\section{Experiments and Results}\label{experiments_results}

Below we list the implementation details of our validation experiment and present the results to demonstrate the effectiveness of our proposed methods. We also report various ablation studies to shed light on the effects of various design decisions.

\subsection{Training details}
We split the dataset based on its size and nature. The SLP dataset consists of 102 subjects and 4590 static posture pressure images, providing sufficient samples for training, evaluation, and testing on unseen subjects. Thus we divide the dataset into ten bins by weight range and randomly select two subjects from each bin for the evaluation and test dataset. Finally, the sample ratio of training, evaluating, and testing sets is 84:10:8. For our static posture dataset containing 10 subjects, we adopted two splitting strategies with 5-fold cross-validation, namely the random split and the Leave-One-Subject-Out~(LOSO) strategy, abbreviated as Ours(Random) and Ours(LOSO), respectively.

Previous work~\cite{sundholm2014smart} shows that image upsampling using bilinear interpolation provides better representation. We thus upsample every pressure image from our dataset by 3 and then smooth it with a 5 × 5 Gaussian filter. Eventually, we pad all images with zeros~(zero-padding) until the final image size is $192\times192$, the same as in the SLP dataset. 

The experiments are all implemented in the PyTorch framework, and run on a Linux server with four NVIDIA RTX3090 GPUs. We choose the Adam optimizer and set the initial learning rate $3\times 10^{-4}$ for the SLP dataset, $5\times 10^{-4}$ for Ours~(LOSO), and $2\times 10^{-4}$ for Ours~(Random) datasets, with a decay rate of 0.25 after every 5 epochs. The warm-up trick is also implemented to accelerate the training procedure. The batch size is 16 for all three datasets. The weight parameter $\lambda$ is set to 0.25. Data augmentation including flip, rotation, and shift is used for model training.

\renewcommand\arraystretch{1.3}
\begin{table*}[t] \scriptsize
\vspace{-0.5cm}
\caption{MassNet's performance Compared with those of the state-of-the-art methods}
\label{tab:overall results}
\centering
\resizebox{\textwidth}{!}{
\begin{tabular}{lllllllll} 
\cline{1-8}
\multicolumn{1}{c}{\multirow{2}{*}{Method}}                    & \multicolumn{1}{c}{\multirow{2}{*}{Parameters}} & \multicolumn{2}{c}{SLP}                                      & \multicolumn{2}{c}{Ours~(LOSO)}                               & \multicolumn{2}{c}{Ours~(Random)}                                                     &   \\ 
\cline{3-8}
\multicolumn{1}{c}{}                                           & \multicolumn{1}{c}{}                            & \multicolumn{1}{c}{MAE~($kg$)} & \multicolumn{1}{c}{MAPE~(\%)} & \multicolumn{1}{c}~{MAE($kg$)} & \multicolumn{1}{c}{MAPE~(\%)} & \multicolumn{1}{c}{MAE~($kg$)} & \multicolumn{1}{c}{MAPE~(\%)}  &   \\ 
\cline{1-8}
\multicolumn{1}{l|}{Linear fitting}& \multicolumn{1}{c|}{2}                          & \multicolumn{1}{c}{$10.45\pm7.00$}        & \multicolumn{1}{c|}{$16.07\pm10.91$}        & \multicolumn{1}{c}{$11.13\pm5.85$}        & \multicolumn{1}{c|}{$19.70\pm9.91$}        & \multicolumn{1}{c}{$9.13\pm5.51$}        & \multicolumn{1}{c}{$15.90\pm9.29$}                &   \\
\multicolumn{1}{l|}{Davoodnia et al.\cite{davoodnia2020deep}} & \multicolumn{1}{c|}{$0.17$M}                          & \multicolumn{1}{c}{$14.10\pm8.97$}        & \multicolumn{1}{c|}{$20.29\pm12.49$}        & \multicolumn{1}{c}{$12.46\pm8.51$}        & \multicolumn{1}{c|}{$22.90\pm16.59$}        & \multicolumn{1}{c}{$10.66\pm7.67$}        & \multicolumn{1}{c}{$19.53\pm14.80$}        &   \\
\multicolumn{1}{l|}{ResNet18}                                  & \multicolumn{1}{c|}{$11.17$M}                          & \multicolumn{1}{c}{$5.42\pm3.64$}        & \multicolumn{1}{c|}{$8.12\pm5.24$}        & \multicolumn{1}{c}{$8.46\pm6.40$}        & \multicolumn{1}{c|}{$15.56\pm13.09$}        & \multicolumn{1}{c}{$1.97\pm2.28$}        & \multicolumn{1}{c}{$3.46\pm3.94$}        &    \\
\multicolumn{1}{l|}{MobileNetV2}                               & \multicolumn{1}{c|}{$2.22$M}                          & \multicolumn{1}{c}{$6.36\pm5.08$}        & \multicolumn{1}{c|}{$9.53\pm7.42$}        & \multicolumn{1}{c}{$8.32\pm5.77$}        & \multicolumn{1}{c|}{$15.01\pm10.63$}        & \multicolumn{1}{c}{$2.40\pm2.56$}        & \multicolumn{1}{c}{$4.09\pm4.22$}        &    \\ 
\cline{1-8}
\multicolumn{1}{l|}{MassNet}                                   & \multicolumn{1}{c|}{$1.80$M-$3.47$M}                          & \multicolumn{1}{c}{\bm{$4.59\pm3.81$}}        & \multicolumn{1}{c|}{\bm{$6.59\pm5.06$}}        & \multicolumn{1}{c}{\bm{$4.86\pm3.82$}}        & \multicolumn{1}{c|}{\bm{$8.48\pm6.31$}}        & \multicolumn{1}{c}{\bm{$1.50\pm1.34$}}        & \multicolumn{1}{c}{\bm{$2.67\pm2.45$}}        &    \\ 
\cline{1-8}
\end{tabular}}
\end{table*}

\renewcommand\arraystretch{1.3}
\begin{table*}[t] \scriptsize
\vspace{-0.2cm}
\centering 
\caption{MassNet's performance with different numbers of branches}
\label{tab:ablation_branch}
\resizebox{\textwidth}{!}{
\begin{tabular}{lllllllll} 
\cline{1-7}
\multirow{2}{*}{Branch}          & \multicolumn{2}{c}{SLP}                                      & \multicolumn{2}{c}{Ours~(LOSO)}                               & \multicolumn{2}{c}{Ours~(Random)}                                                         &  &   \\ 
\cline{2-7}
& \multicolumn{1}{c}{MAE~(kg)} & \multicolumn{1}{c}{MAPE~(\%)} & \multicolumn{1}{c}{MAE~(kg)} & \multicolumn{1}{c}{MAPE~(\%)} & \multicolumn{1}{c}{MAE~(kg)} & \multicolumn{1}{c}{MAPE~(\%)}  &  &   \\ 
\cline{1-7}
\multicolumn{1}{l|}{Joint Branch~(JT) only}          & \multicolumn{1}{c}{$14.83\pm10.33$}        & \multicolumn{1}{c|}{$19.64\pm11.85$}        & \multicolumn{1}{c}{$10.40\pm7.80$}        & \multicolumn{1}{c|}{$18.49\pm13.64$}        & \multicolumn{1}{c}{$2.41\pm2.61$}        & \multicolumn{1}{c}{$4.10\pm4.27$}                 &  &   \\
\multicolumn{1}{l|}{Mass branch~(Mass) only}  & \multicolumn{1}{c}{$5.42\pm4.44$}        & \multicolumn{1}{c|}{$7.72\pm5.49$}        & \multicolumn{1}{c}{$5.03\pm3.89$}        & \multicolumn{1}{c|}{$8.79\pm6.48$}        & \multicolumn{1}{c}{\bm{$1.32\pm1.34$}}        & \multicolumn{1}{c}{\bm{$2.34\pm2.44$}}                  &  &   \\
\multicolumn{1}{l|}{JT+Mass} & \multicolumn{1}{c}{\bm{$4.59\pm3.81$}}        & \multicolumn{1}{c|}{\bm{$6.59\pm5.06$}}        & \multicolumn{1}{c}{\bm{$4.86\pm3.82$}}        & \multicolumn{1}{c|}{\bm{$8.48\pm6.31$}}        & \multicolumn{1}{c}{$1.50\pm1.34$}        & \multicolumn{1}{c}{$2.67\pm2.45$}                &  & \\ 
\cline{1-7}
\end{tabular}}
\end{table*}

\subsection{Evaluation Methodology}

We compare our model with the following four baselines: (1) Linear Fitting: Shall the pressure and sensor output have a deterministic and linear correlation, and the overall pressure exerted on the bedsheet be equal to the product of body weight and gravity, a liner fitting would perfectly deduce weight from the pressure image.  We thus use the linear fitting to construct the relationship between the value sum of each frame and its label in the training set, then make predictions in the testing set. (2) Feature Extraction: we extract the features proposed by~\cite{davoodnia2020deep} and train 
datasets with the same DNN network. (3) ResNet~\cite{he2016deep}: one of the most famed vision-based networks that has shown excellence in many downstream tasks and also proved productively on RGB-based weight estimation task in~\cite{jin2022estimating}. We choose RestNet18 finally owing to our data size.  (4) MobileNetV2~\cite{sandler2018mobilenetv2}: a noted lightweight network that holds high performance with less consumption. We adopt it for its recognized capabilities and close number of parameters.

To compare the performance of the proposed method with other baselines, two metrics are adopted: the mean absolute error~(MAE) and the mean absolute percentage error~(MAPE). 

\subsection{Results} \label{overall}
Table I exhibits the quantitative comparison of accuracy and model complexity on different datasets. All results are represented as $mean\pm std$. 

Our proposed method~(MassNet) achieves minimal errors compared to all other baseline algorithms and reaches only $1.5kg$ MAE loss on Ours~(Random), showing marked weight prediction capacities. Meanwhile, MassNet outperforms the state-of-the-art methods on SLP and Ours~(LOSO) by a significant 15.31\% and 42.55\% decrease in MAE loss, respectively. ResNet18 also performs well on the SLP dataset, showing strong interpretability abilities with 11.17 million parameters, and thus overfits and works under-performingly on small datasets such as ours~(LOSO) dataset. Moreover, all three deep neural networks perform superiorly on Ours~(Random), with less than $2.5kg$ prediction errors and 5\% relative errors. Considering the sensor mat's non-ideal characteristics aforementioned in Section~\ref{sec:datadescrp} and their limited spatial resolutions, this result can be regarded as a surprise. 

Also because of the non-ideal characteristics, the Linear Fitting method shows limited prediction precision on both the SLP and our datasets, just as expected. The method in~\cite{davoodnia2020deep} also failed to precisely predict, probably because the onefold 14 statistical features are not enough to describe the correlation between pressure distribution and body weights.  


\subsection{Ablation Study}
To evaluate the effectiveness of MassNet's individual components, ablation studies are conducted.

\subsubsection{Contributions of the two branches}\label{ablation_study_branch}

To gain insight into the roles of the two branches in MassNet with refined contrastive loss, we have conducted various ablation experiments to analyze each branch's performance on all datasets. All these experiments were conducted on the same training and testing sets. The results are shown in Table~\ref{tab:ablation_branch}. Despite joint positions reflecting human pose information, the single joint branch performs poorly on three environment setups. Deep features extracted by the Mass branch bring nearly 50\% promotion compared to the single joint branch and reach the best result on Ours~(random), with only $1.32kg$ MAE loss. Furthermore, MassNet with dual branches supplies incremental precision improvement compared to the single branches, with the MAE loss decreasing by 15.31\% on the SLP dataset and 3.38\% on Ours~(LOSO). However, the prediction results with dual-branch architecture are worse than the single body weight branch on Ours~(Random). In future works, more care shall be given to how to utilize the joint vectors.

\subsubsection{Contributions of refined contrastive loss}\label{ablation_study_conloss}

\begin{figure}[t]
\vspace{-0.5cm}
\setlength{\abovecaptionskip}{-0.15cm}
\centerline{\includegraphics[width=0.5\textwidth]{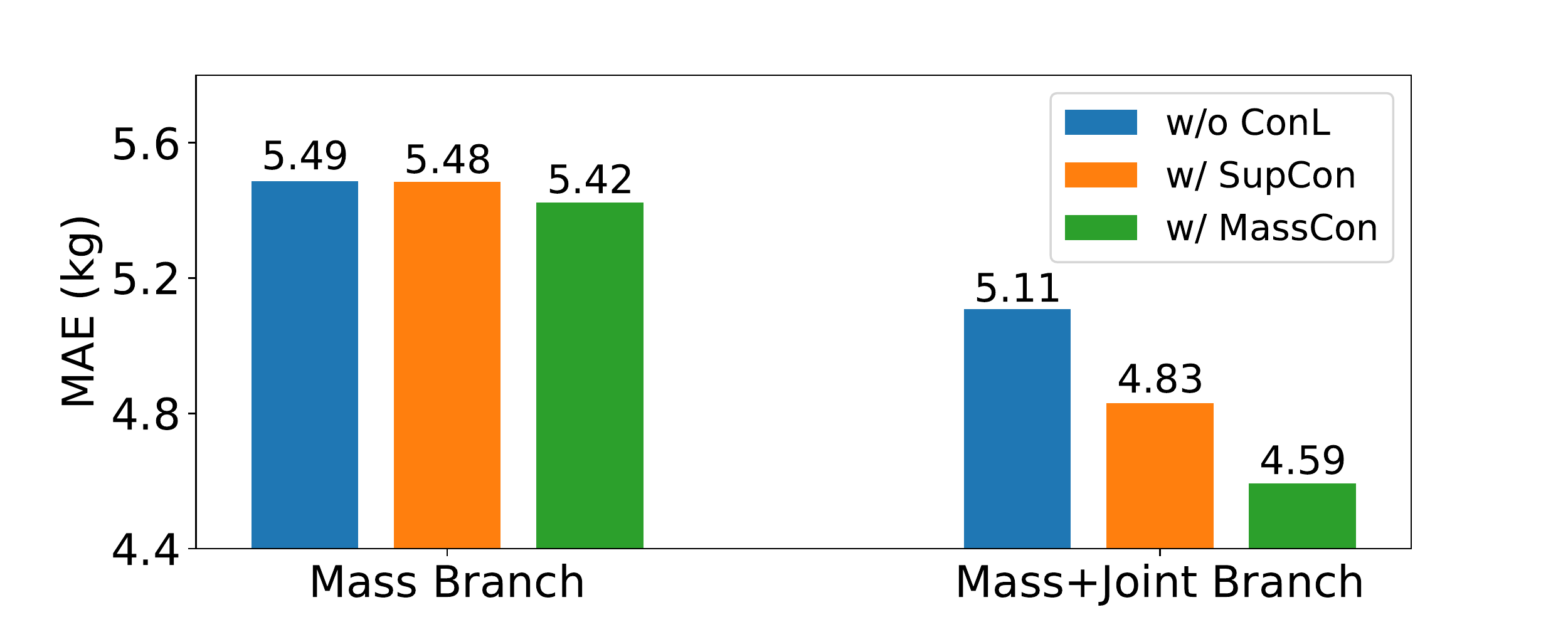}}
\caption{Performances with/without the contrastive loss module}
\label{fig:ablation_con_l}
\end{figure}

We test MassNet with and without contrastive loss~(abbreviated as ConL), which is a significant component of this study to learn the posture-invariant representations. The supervised contrastive loss~\cite{khosla2020supervised}~(abbr. SupCon) is also implemented as a comparison parameter to evaluate our refined contrastive loss(abbr. MassCon). The experiment is conducted only on the SLP dataset because it contains the most subjects~(102), providing enough training samples for contrastive learning. Fig.~\ref{fig:ablation_con_l} shows that MassCon archives the best performance both in the single-branch and double-branch MassNet architectures by reducing the overall MAE by $0.07kg$ and $0.52kg$ compared to the network without ConL, while SupCon only brings $0.01kg$ and $0.28kg$ prediction errors decreases. This result demonstrates that the auxiliary contrastive loss successfully helps extract the subject-variant deep attributes, and the penalty factor in MassCon guides the network to digest the weight-related attributes. Moreover, the result demonstrates the bottleneck that the contrastive module faces due to data size and posture diversity and proves that the auxiliary posture information can support the final regression module to interpret the posture-invariant features and improve the prediction accuracy. We plan to continue investigating this association in future works.

\begin{figure}[t]
\vspace{-0.55cm}
\setlength{\abovecaptionskip}{-0.05cm}
\centerline{\includegraphics[width=0.5\textwidth, height=4cm]{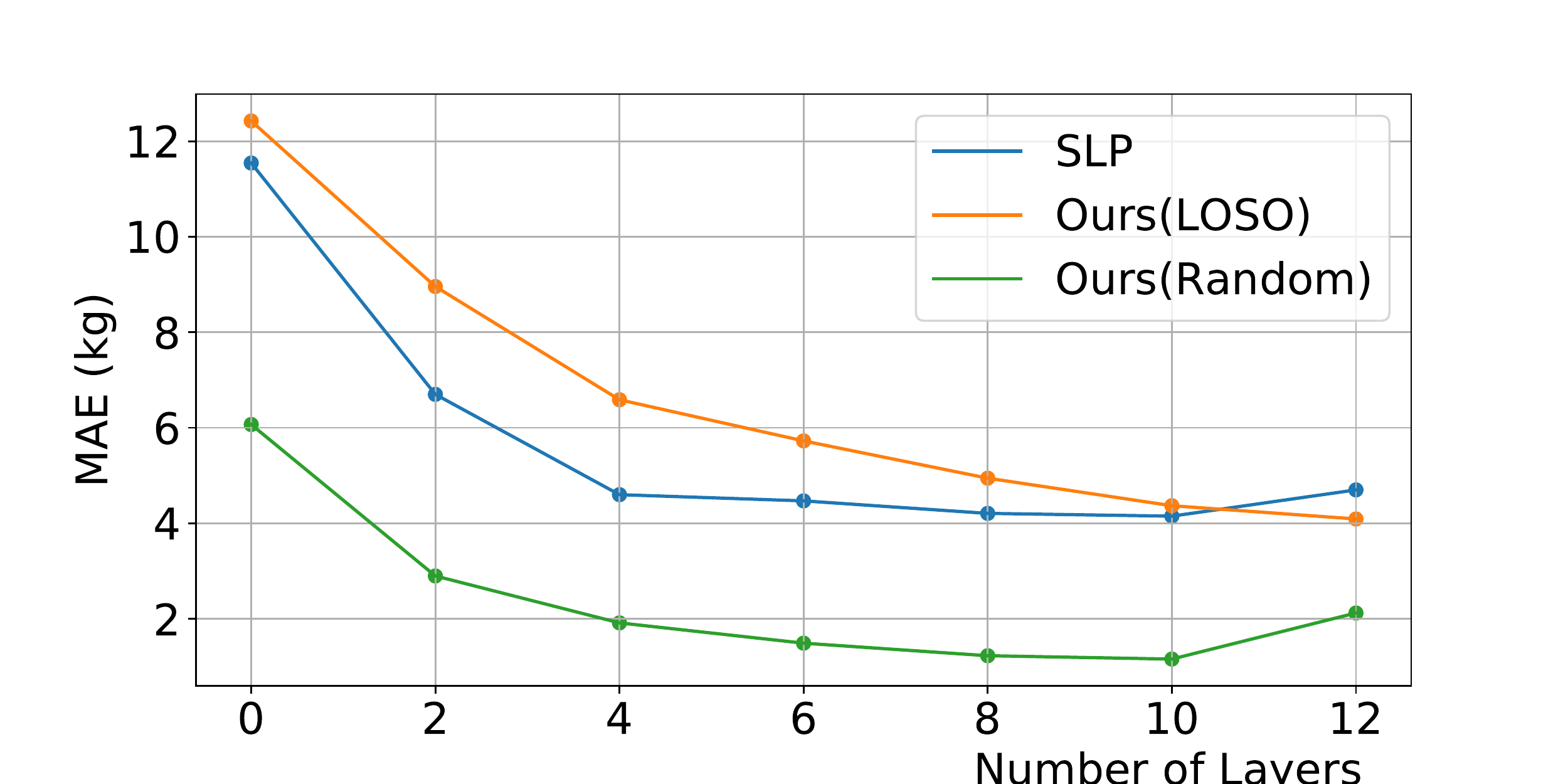}}
\caption{Performances with different numbers of sensing layers}
\label{fig:ablation_block}
\end{figure}

\subsubsection{Contributions of sensing layers}

Deeper networks bring stronger inference and reasoning abilities, but more training samples and resources are required. Hence, we conduct a scan to find out the suitable number for the stacked sensing layers. Fig.~\ref{fig:ablation_block} presents the prediction errors of our MassNet with the number of sensing layers increasing from 0 to 12. The MAE drops rapidly at the outset and then levels off and rises due to overfitting with the sensing layers up to 10 for both the SLP dataset and Ours. Thus, to balance the precision and consumption, stacked layer numbers are carefully picked for each dataset and are set to 4 for SLP, 8 for Ours~(LOSO), and 6 for Ours~(random) as the optimal parameters. 

\subsubsection{Performance on different postures}

\begin{figure}[]
\centering
\setlength{\abovecaptionskip}{-0.05cm}
\setlength{\subfigcapskip}{-0.1cm}
\vspace{-0.2cm}
   \subfigure[the SLP dataset]{\includegraphics[width=0.45\textwidth]{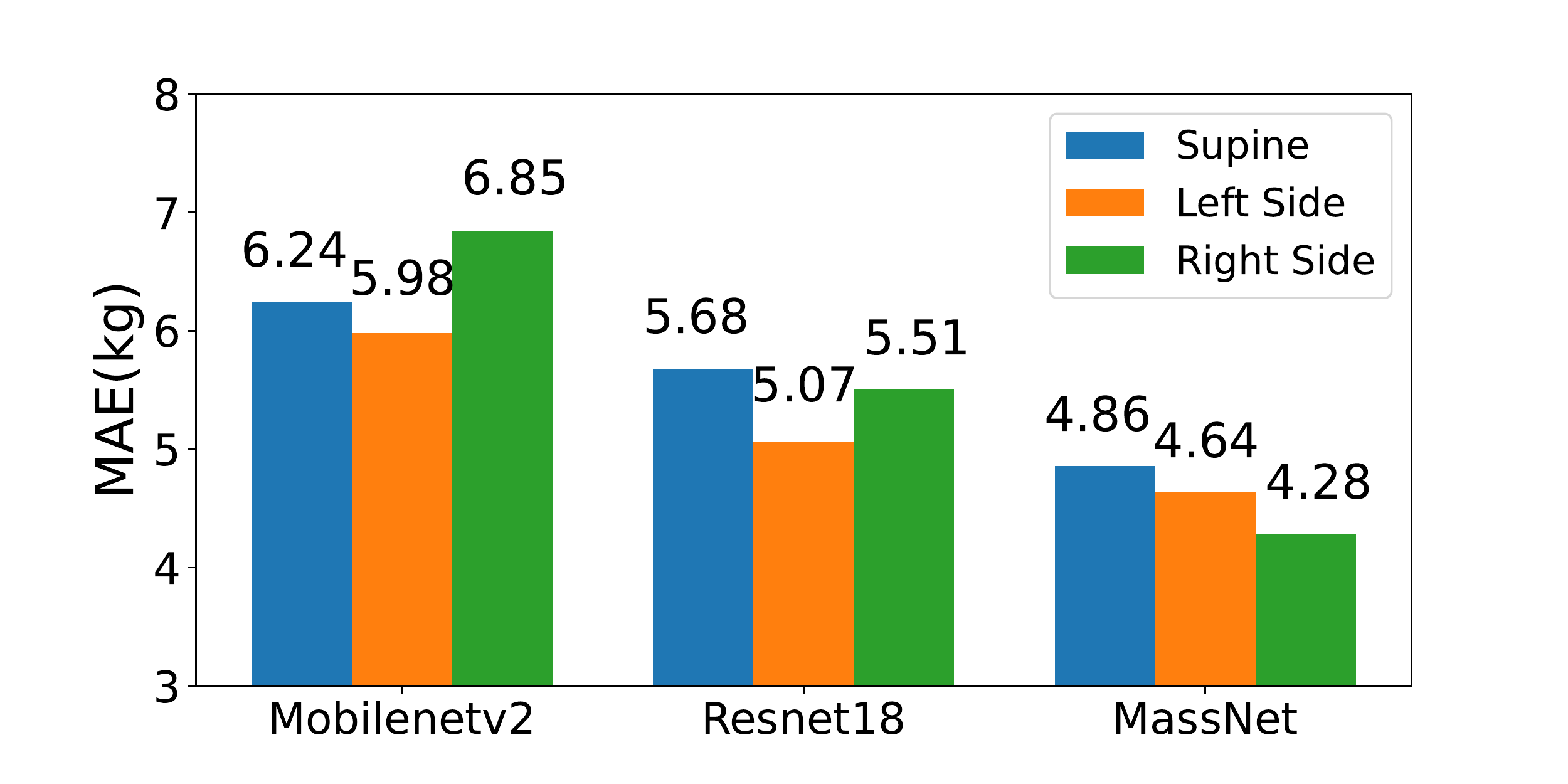}\label{fig:ablation_posture_slp}} \vskip -0.05cm
   \subfigure[Ours~(LOSO) dataset]{\includegraphics[width=0.45\textwidth]{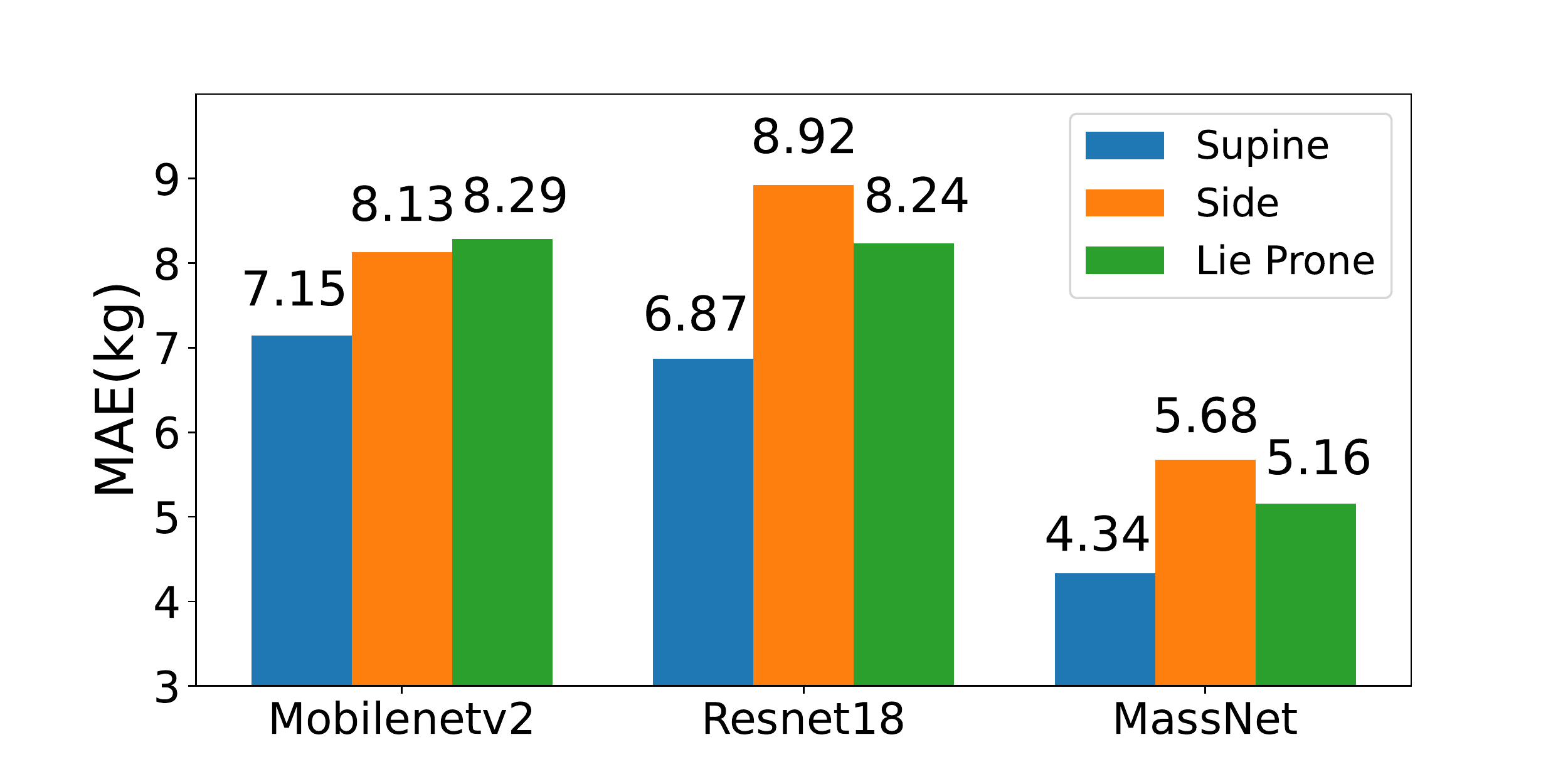}\label{fig:ablation_posture_ours}}
\caption{MassNet's performance on different postures}
\label{fig:ablation_posture}
\end{figure}

\begin{figure}[ht]
\vspace{-0.2cm}
\setlength{\abovecaptionskip}{-0.01cm}
\centering
   \subfigure[Supine I]{\includegraphics[width=0.15\textwidth, height=4cm]{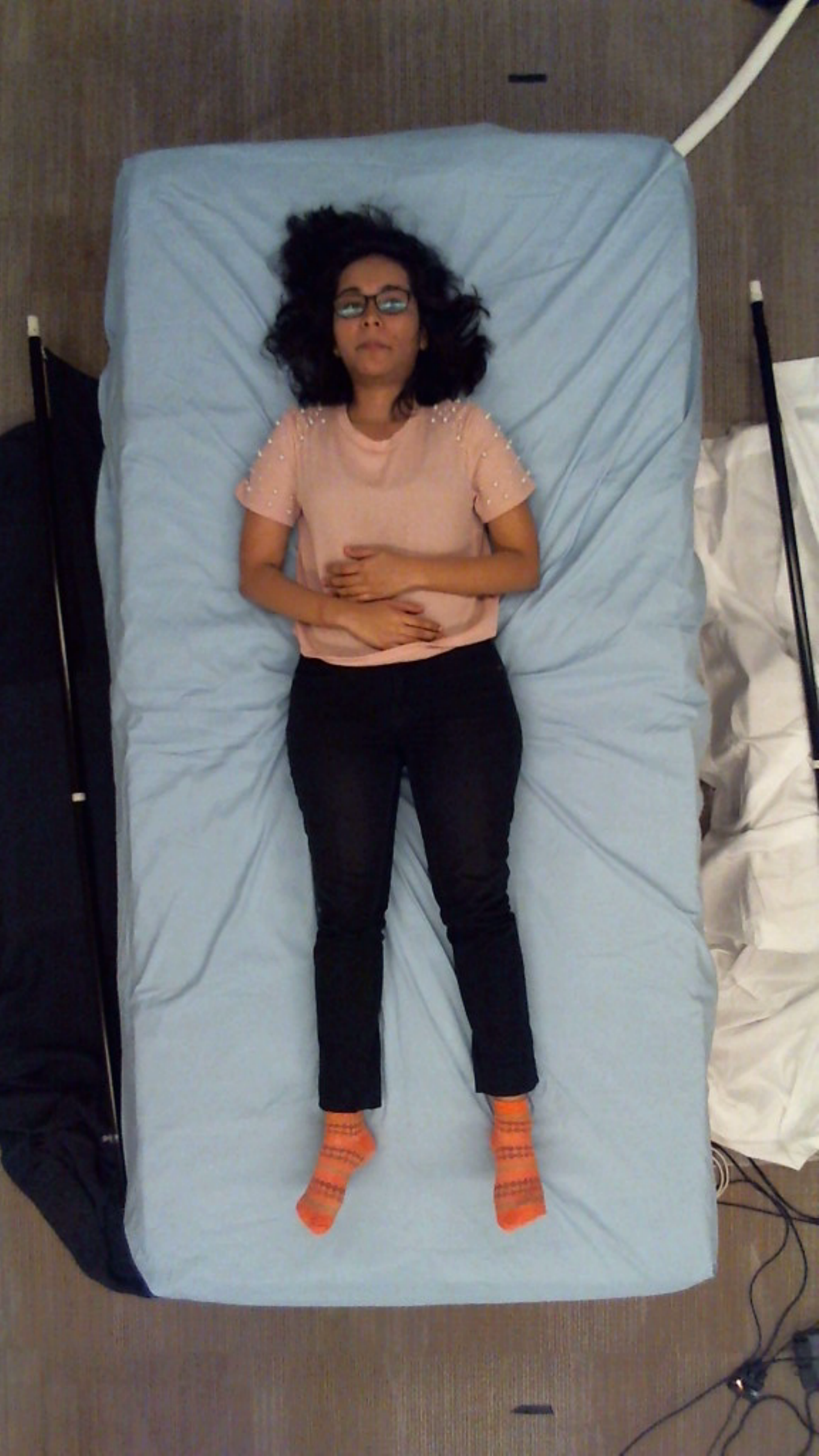}\label{fig:supineI}}
   \subfigure[Supine II]{\includegraphics[width=0.15\textwidth, height=4cm]{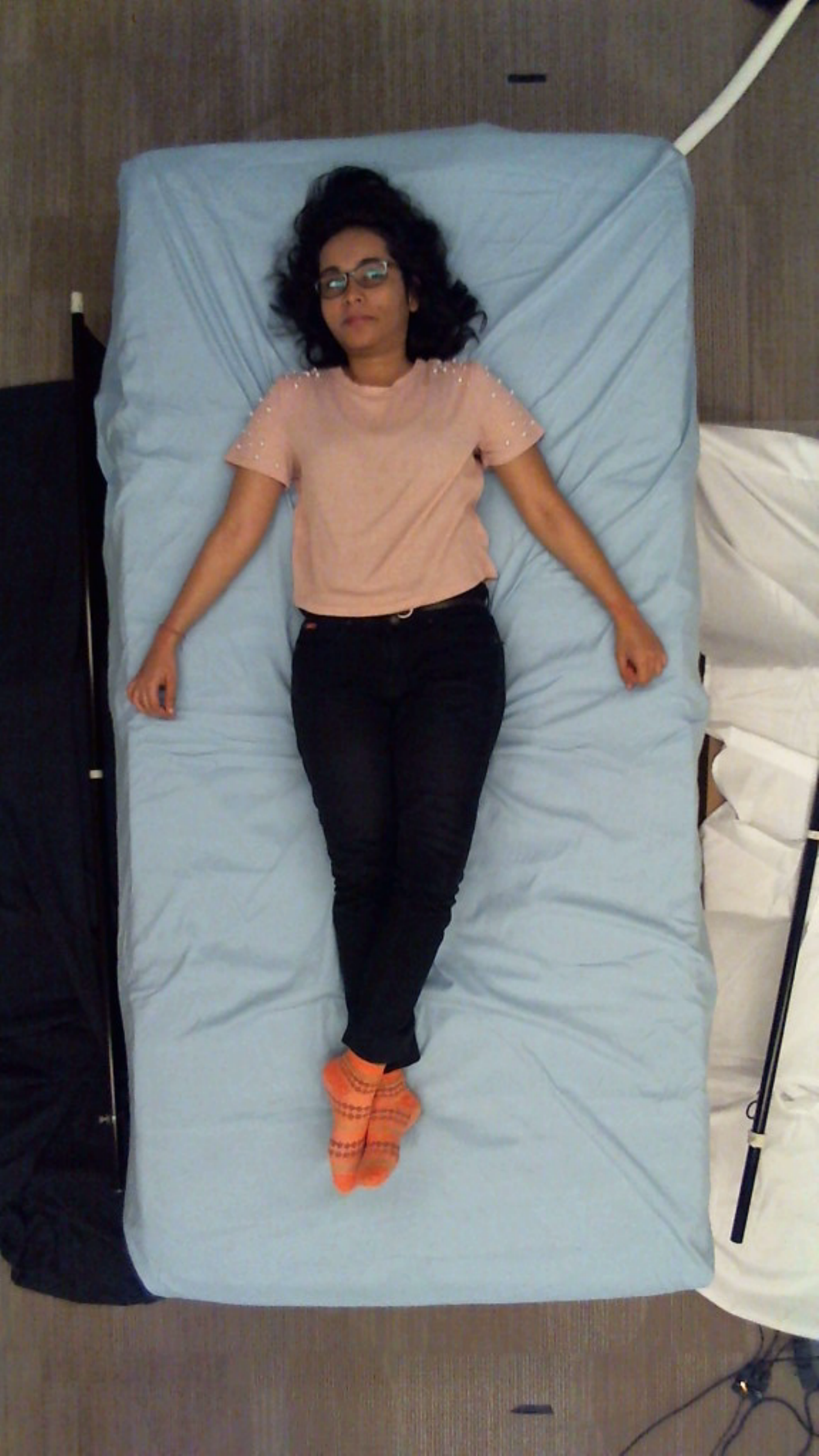}\label{fig:supineII}}
   \subfigure[Supine III]{\includegraphics[width=0.15\textwidth, height=4cm]{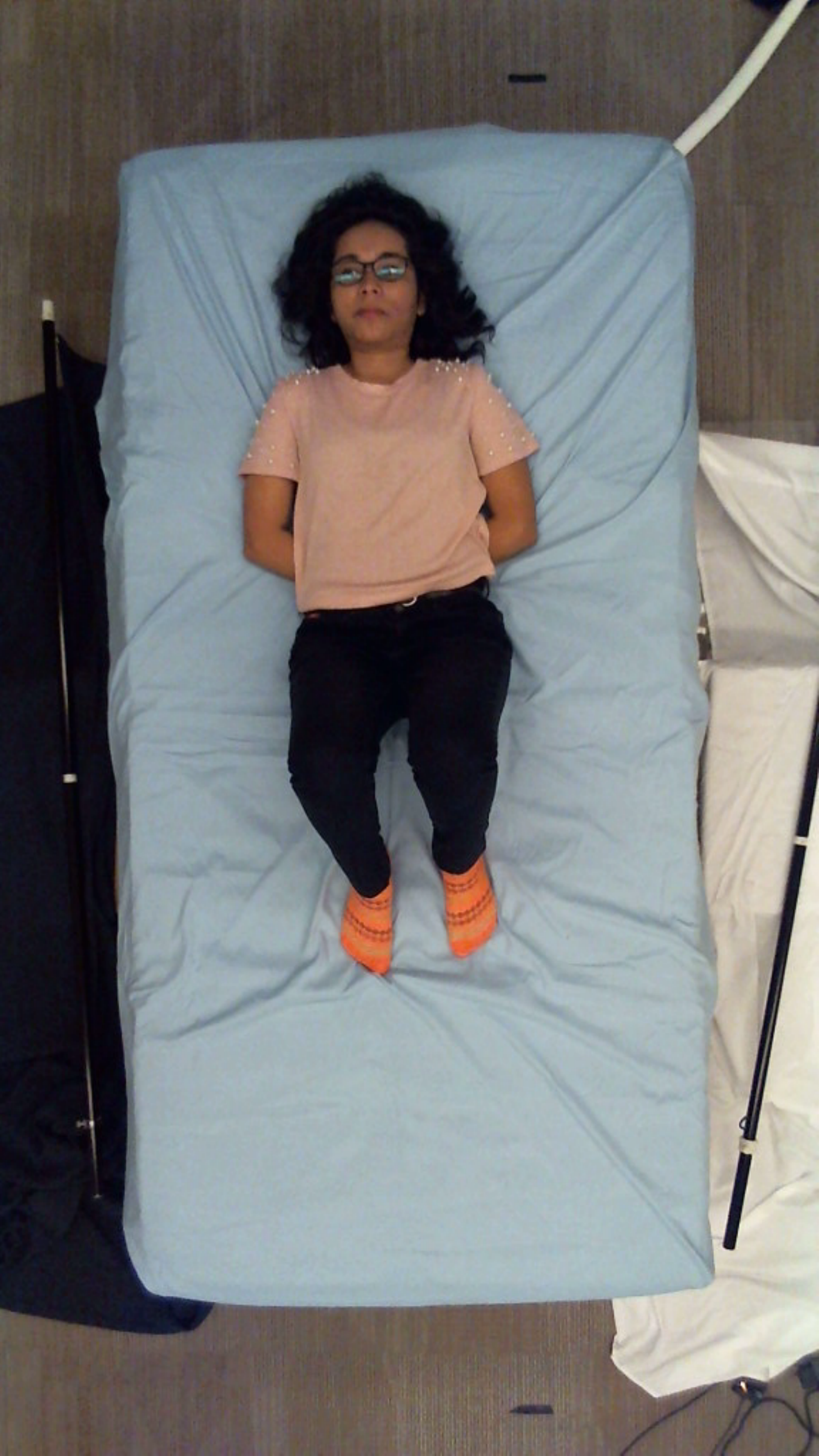}\label{fig:supineIII}}
    \subfigure[Predicted: $41.41kg$]{\includegraphics[width=0.15\textwidth, height=4cm]{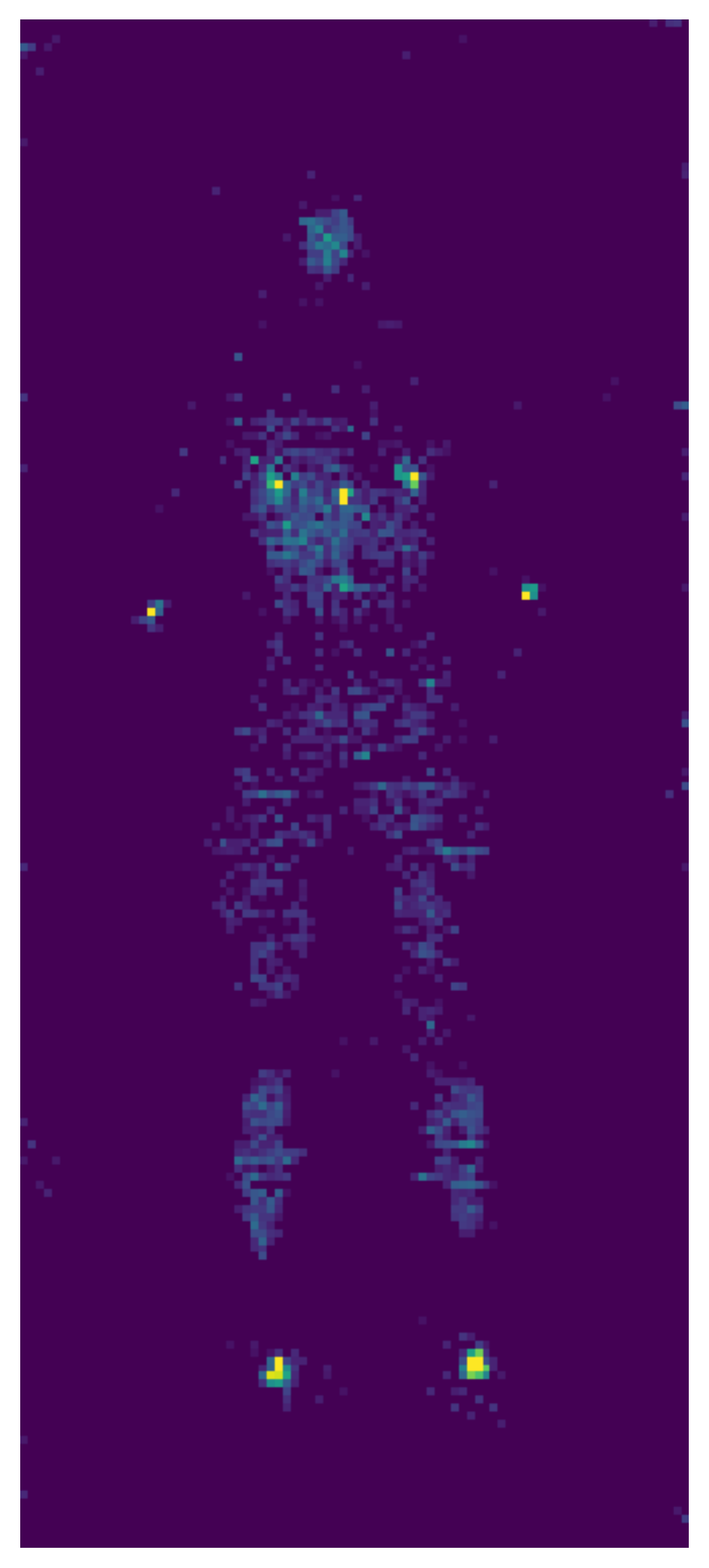}\label{fig:supineIPImg}}
   \subfigure[Predicted: $48.59kg$]{\includegraphics[width=0.15\textwidth, height=4cm]{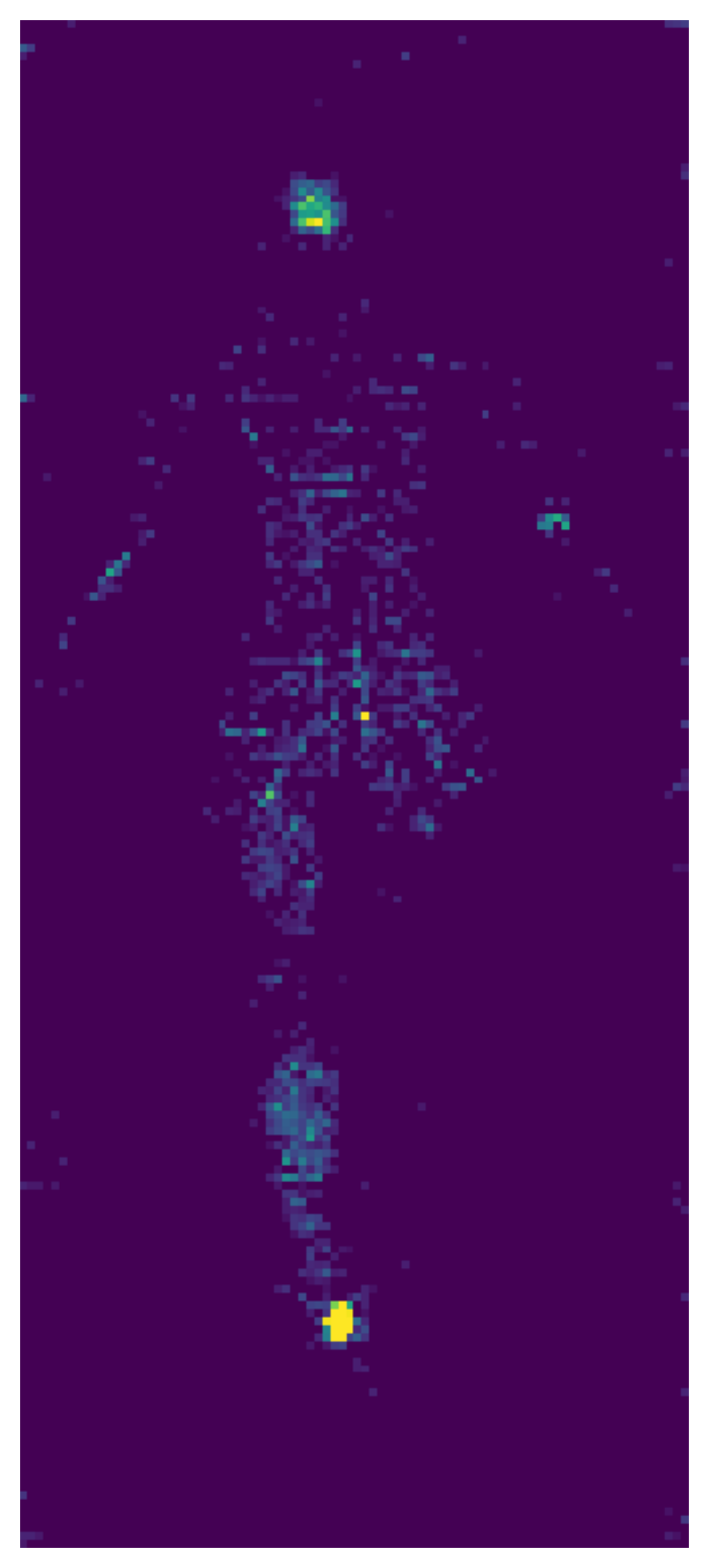}\label{fig:supineIIPImg}}
   \subfigure[Predicted: $52.43kg$]{\includegraphics[width=0.15\textwidth, height=4cm]{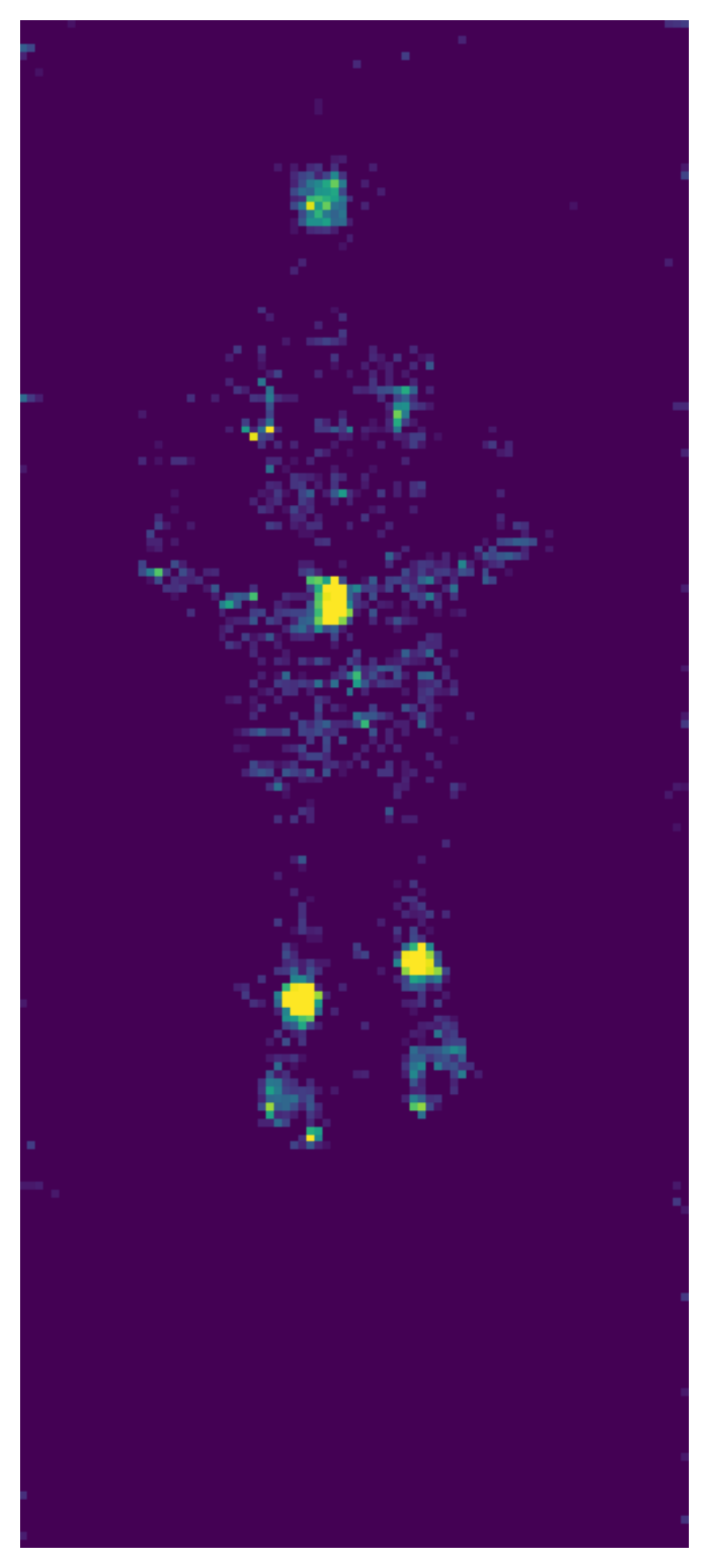}\label{fig:supineIIIPImg}}
\caption{MassNet predictions on a subject from SLP~($48kg$ as Ground Truth)}
\label{fig:ablation_posture_subject}
\end{figure}

Different postures might assert different difficulties for weight predictions. We discuss their influence below. Fig.~\ref{fig:ablation_posture_slp} shows the three deep network performances on three posture types in the SLP dataset. Intuitively, models shall perform the best on the pressure images of supine postures because the body contour and pressure features could be fully displayed. The results however suggest the exact opposite. The reason behind might be that the hand positions and body huddles confuse the network, as demonstrated in Fig.~\ref{fig:ablation_posture_subject}. Putting the hands on the chest decreases the contact area, resulting in a small prediction, while body huddles may mislead the network in a fatter body shape, resulting in a higher weight prediction. Thus, we instructed subjects' movements to avoid curled-up or uncomfortable positions during our data collection, and pressure images of supine positions achieve as expected the minimal MAE compared to the other postures~(Fig.~\ref{fig:ablation_posture_ours}).

\subsubsection{The joint positions' information source} \label{joint_accuracy}

As mentioned in Section~\ref{chap:joint_fea_ex}, pressure images' precise 2D/3D joint location annotations normally rely on vision-based pose estimation networks applied onto their aligned RGB images, severely restricting the application scenarios. While sleeping in bed is normally linked with strict privacy and dark illumination, the vision-based pose estimation would encounter troubles. HRNet~\cite{sun2019deep} is a state-of-the-art pose estimation method and has been  proven to be effective on pressure images in ~\cite{liu2020simultaneously}. We use the same training subset for MassNet to train the HRNet, and get an average of 5 to 9 pixel errors on the test sets, as shown in Table~\ref{tab:ablation_joint}. The computed joint vectors and their corresponding pressure images are then fed to MassNet for the body weight estimation task. Table~\ref{tab:ablation_joint} shows the maximal precision loss on the SLP dataset is 0.47, about a 10\% plus compared to the results with precise joint positions, and the precision change can nearly be ignored on the other datasets. Furthermore, the outcomes outperform the model with the only mass branch on all datasets except Ours~(Random). This study proves the possibility to apply our MassNet to illumination-sensitive and privacy-concerning environments.

\renewcommand\arraystretch{1.5}
\begin{table}[t] \scriptsize
\vspace{-0.5cm}
\caption{Performance with accurate or computed joint positions}
\label{tab:ablation_joint}\centering
\resizebox{0.5\textwidth}{!}{
\begin{tabular}{llllll} 
\cline{1-4}
\multicolumn{1}{l|}{}                                      & SLP & Ours~(LOSO) & Ours~(Random)  &  &   \\ 
\cline{1-4}
\cline{1-4}
\multicolumn{1}{l|}{Joint Prediction Error(L1 Loss)}        & \multicolumn{1}{c}{$7.45$}   & \multicolumn{1}{c}{$7.72$}          & \multicolumn{1}{c}{$5.25$}                       &  &   \\
\multicolumn{1}{l|}{MAE: without joints~($kg$)}        & \multicolumn{1}{c}{$5.42$}   & \multicolumn{1}{c}{$5.03$}          & \multicolumn{1}{c}{$1.32$}                         &  &   \\
\multicolumn{1}{l|}{MAE: joints from HRNet~($kg$)}        & \multicolumn{1}{c}{$5.06$}   & \multicolumn{1}{c}{$4.90$}          & \multicolumn{1}{c}{$1.49$}                       &  &   \\
\multicolumn{1}{l|}{MAE: joints from ground truth~($kg$)} & \multicolumn{1}{c}{$4.59$}   & \multicolumn{1}{c}{$4.86$}          & \multicolumn{1}{c}{$1.50$}                        &  &   \\
\multicolumn{1}{l|}{Performance loss}                      & \multicolumn{1}{c}{$-0.47$}   & \multicolumn{1}{c}{$-0.04$}          & \multicolumn{1}{c}{$-0.01$}                       &  &   \\ 
\cline{1-4}
\cline{1-4}
\end{tabular}}
\end{table}

\subsubsection{Performance on time series pressure data} 

\begin{figure}[t]
\vspace{-0.5cm}
\setlength{\abovecaptionskip}{-0.01cm}
\centering
   \subfigure[The pressure value sum curve to show the subject's movement]{\includegraphics[width=0.5\textwidth, height=4cm]{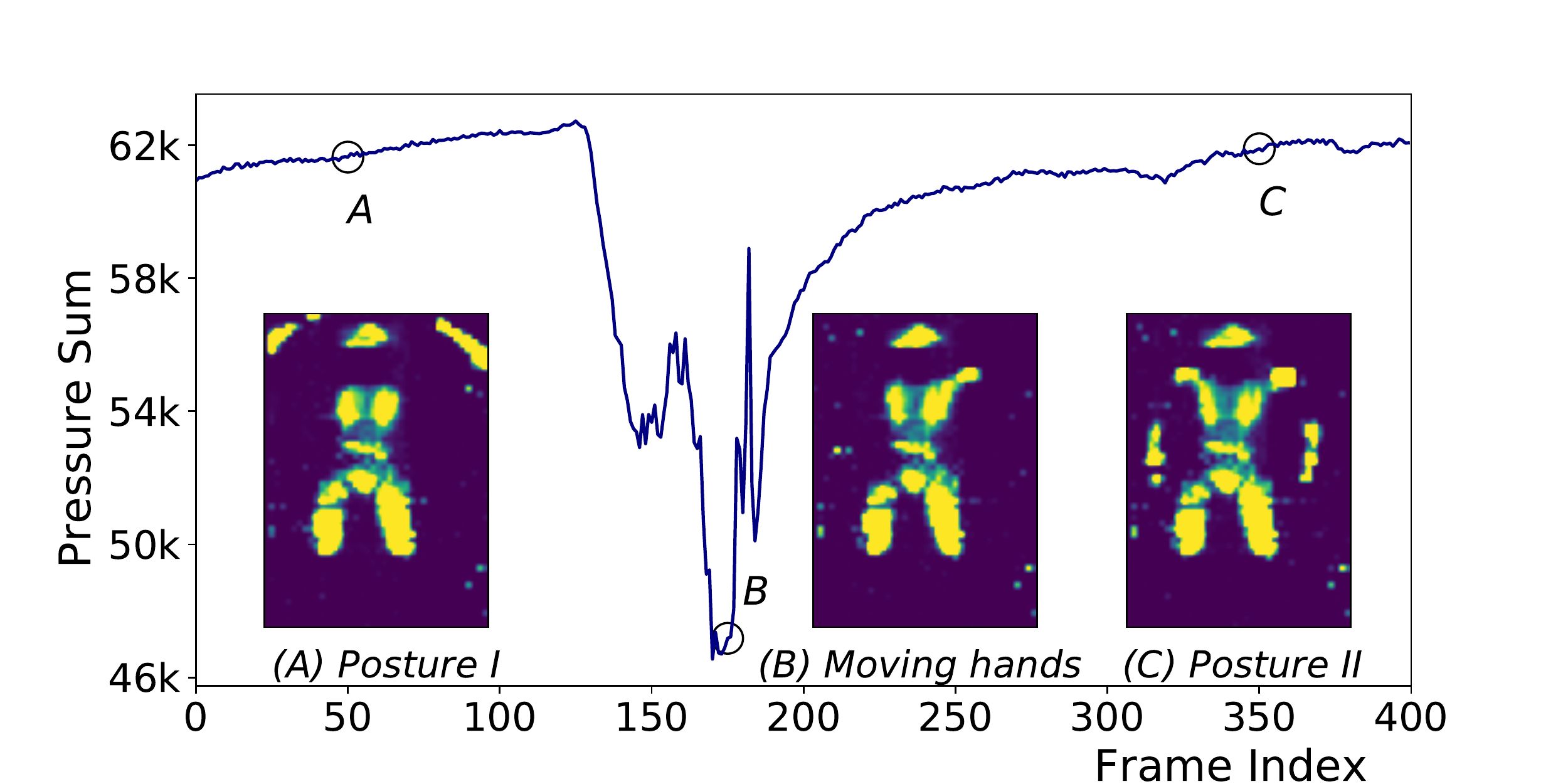}\label{fig:ablation_series_pre_sum}} \vskip -0.15cm
   \subfigure[The model's prediction curve and the ground truth]{\includegraphics[width=0.5\textwidth, height=4cm]{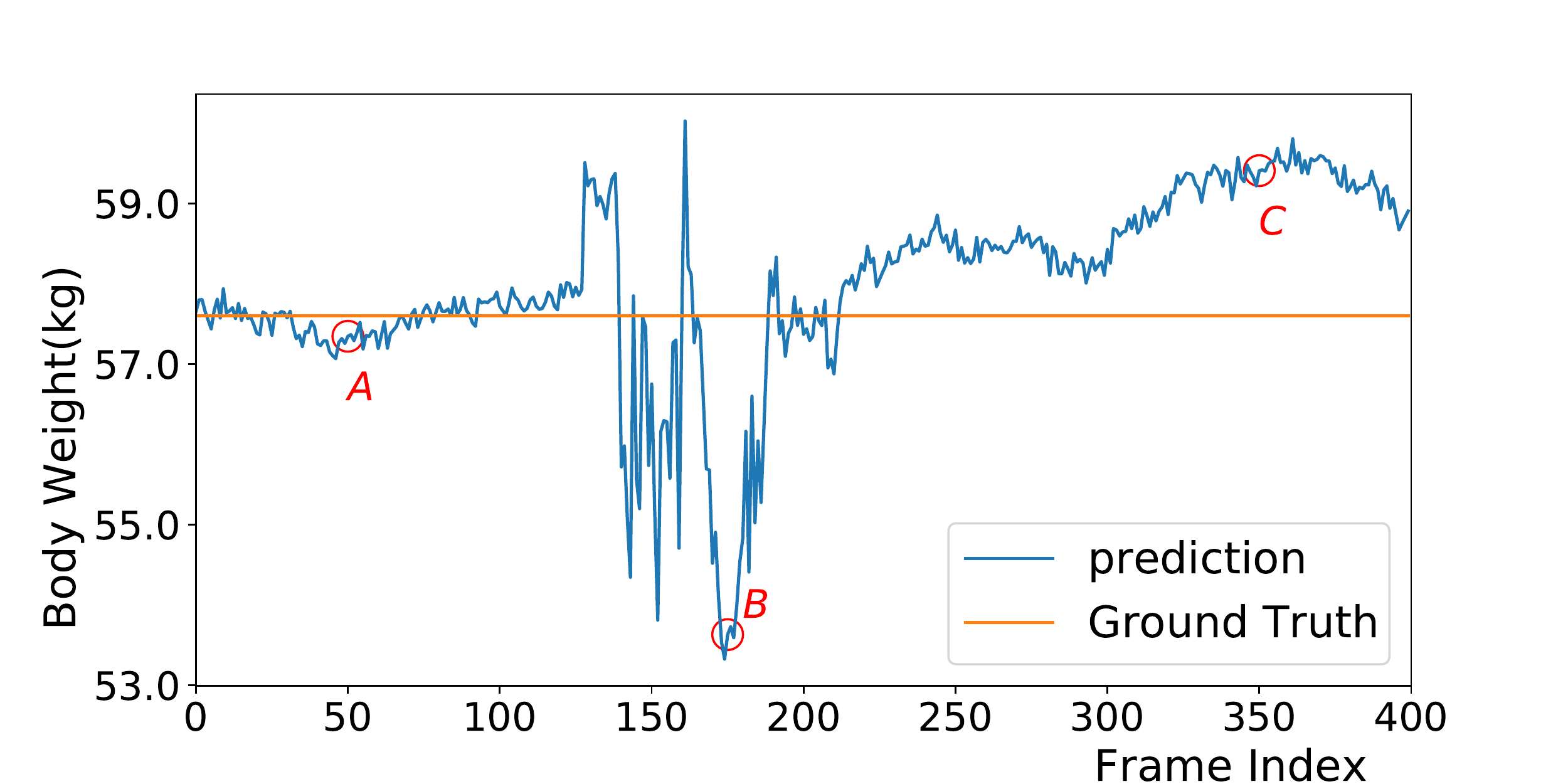}\label{fig:ablation_series_weight}}
\caption{Performance on time series data}
\label{fig:ablation_series_data}
\end{figure}

To verify MassNet's stability and robustness in the long-term weight monitoring task, we train it with the static posture images, then apply it onto the whole dynamic dataset, which contains more than 36000 frames. Given that the subject's movements can cause perturbations in the pressure values, a gradient-based active frame selection method is implemented to split the whole time-series dataset into static and active parts. MassNet achieves a $4.70(\pm0.90)kg$ MAE in the static part and $9.58(\pm3.82)kg$ in the active part. Fig.~\ref{fig:ablation_series_data} gives a glimpse of our method's performance. When a subject was asked to move the hands from the head to the ventral side, as shown in Fig.~\ref{fig:ablation_series_pre_sum}, the predicted weight drops simultaneously as the contact area decreases. The predicted body weight returns quickly to a stable value after the action is finished. 

\section{Discussions and Conclusions}\label{conclusions}

We have proposed a dual-branch neural network, the MassNet, that can extract body weight from a single pressure image. This model outperforms the baseline models, namely the linear fitting, feature extraction, and two state-of-the-art neural networks, on both a public dataset with a film sensing mat and a self-created dataset with a pure-textile mat. With ablation studies, we proved the usefulness of the two branches and the contrastive loss module, studied the influence of different postures and movements on the weight estimation task, and proved that the MassNet can also work fully independently without any support from the extra camera(s).

For future works, the MassNet could be further enhanced at least by three methods:1) to give more concrete studies about the network hyperparameters and setups like loss weight $\lambda$, network channels, etc. and elaborate discussion on how they are tuned in practice. (2) to create more branches. For example, another branch of body shape parameters could be added, where the body shape parameters can be again deduced from the pressure image and enhanced by averaging the results from multiple postures. (3) to take the time-series data into consideration. As a person's body weight remains nearly constant in a short sleeping period, the predictions from multiple postures during the whole sleeping period might be merged~(e.g. averaged with their confidence, which is estimated by the postures) to achieve a more accurate result. 

Lastly, the datasets used in this paper are all collected in the laboratory configuration, dissimilar to in-wild environments. Hence the next step would be to push the MassNet into real-life conditions, e.g. considering disturbance and noises from environmental objects like pillows and heavy blankets.


\normalem

\end{document}